\documentclass[twoside,twocolumn]{article}

\usepackage[english]{babel} % Language hyphenation and typographical rules

\usepackage[hmarginratio=1:1,left=12mm, right=12mm,top=20mm,bottom=20mm,columnsep=1cm]{geometry} % Document margins
\usepackage[hang, small,labelfont=bf,up,textfont=it,up]{caption} % Custom captions under/above floats in tables or figures
\usepackage{booktabs} % Horizontal rules in tables

\usepackage{abstract} % Allows abstract customization
\usepackage{titlesec} % Allows customization of titles
\renewcommand\thesection{\Roman{section}} % Roman numerals for the sections
\renewcommand\thesubsection{\roman{subsection}} % roman numerals for subsections
\titleformat{\section}[block]{\large\scshape\centering}{\thesection.}{1em}{} % Change the look of the section titles
\titleformat{\subsection}[block]{\large}{\thesubsection.}{1em}{} % Change the look of the section titles

\usepackage{titling} % Customizing the title section

\usepackage{hyperref} % For hyperlinks in the PDF
\usepackage{color}
\usepackage{graphicx}
\usepackage{caption}
\usepackage{amsmath}
\usepackage{amsfonts}
\usepackage{longtable}
\usepackage{mathtools, cuted}

\usepackage{url}

\usepackage{breakurl}

\pretitle{\begin{center}\Huge\bfseries} % Article title formatting
	\posttitle{\end{center}} % Article title closing formatting
\title{A global model for predicting the arrival of imported dengue infections} % Article title
\author
{Jessica Liebig$^{1}$\thanks{Corresponding author: jess.liebig@csiro.au}, Cassie Jansen$^2$, Dean Paini$^3$, Lauren Gardner$^{1,4,5}$, Raja Jurdak$^{1,6,7}$\\
	\\
	\normalsize{$^{1}$Data61, Commonwealth Scientific and Industrial Research Organisation}\\
	\normalsize{Brisbane, Queensland, Australia}\\
	\normalsize{$^{2}$Communicable Diseases Branch, Department of Health}\\
	\normalsize{Brisbane, Queensland, Australia}\\
	\normalsize{$^{3}$Health \& Biosecurity, Commonwealth Scientific and Industrial Research Organisation}\\
	\normalsize{Canberra, Australian Capital Territory, Australia}\\
	\normalsize{$^{4}$Department of Civil Engineering, Johns Hopkins University}\\
	\normalsize{Baltimore, Maryland, USA}\\
	\normalsize{$^{5}$School of Civil and Environmental Engineering, University of New South Wales}\\
	\normalsize{Sydney, New South Wales, Australia}\\
	\normalsize{$^{6}$School of Electrical Engineering and Computer Science, Queensland University of Technology}\\
	\normalsize{Brisbane, Queensland, Australia}\\
	\normalsize{$^{7}$School of Computer Science and Engineering, University of New South Wales}\\
	\normalsize{Sydney, New South Wales, Australia}\\
}
\date{} % Leave empty to omit a date
\renewcommand{\maketitlehookd}{%
	\begin{abstract}
		\noindent  
		With approximately half of the world's population at risk of contracting dengue, this mosquito-borne disease is of global concern. International travellers significantly contribute to dengue's rapid and large-scale spread by importing the disease from endemic into non-endemic countries. To prevent future outbreaks and dengue from establishing in non-endemic countries, knowledge about the arrival time and location of infected travellers is crucial. We propose a network model that predicts the monthly number of dengue-infected air passengers arriving at any given airport. We consider international air travel volumes to construct weighted networks, representing passenger flows between airports. We further calculate the probability of passengers, who travel through the international air transport network, being infected with dengue. The probability of being infected depends on the destination, duration and timing of travel. Our findings shed light onto dengue importation routes and reveal country-specific reporting rates that have been until now largely unknown. This paper provides important new knowledge about the spreading dynamics of dengue that is highly beneficial for public health authorities to strategically allocate the often limited resources to more efficiently prevent the spread of dengue.
	\end{abstract}
}

\begin{document}
	
	% Print the title
	\maketitle
	
	%----------------------------------------------------------------------------------------
	%	ARTICLE CONTENTS
	%----------------------------------------------------------------------------------------
	
\section*{Introduction}
The well connected structure of the global air transportation network and the steadily increasing volume of international travel has a vast impact on the rapid, large-scale spread of arboviral and other diseases~\cite{Bloom2017,Brockmann2013,Dorigatti2017,Guimera2005,Huang2013,Hufnagel2004,Tatem2012}. A recent example of disease introduction to a novel region is the spread of the Zika virus from Brazil to Europe, the United States and other countries, which prompted the World Health Organisation (WHO) to announce a public health emergency of international concern in early 2016. Investigations confirmed that international viraemic travellers were a major contributing factor to the rapid spread~\cite{Bogoch2016}.

With an estimated 50-100 million symptomatic infections each year~\cite{Bhatt2013,Stanaway2016}, dengue is ranked the most important mosquito-borne disease~\cite{Murray2013,WorldHealthOrganization2013}. The rapid geographic spread is, to a great extent, driven by the increase in international air travel~\cite{Gardner2013,Mackenzie2004}. In addition, dengue is severely under-reported, making it extremely challenging to monitor and prevent the spread of the disease. Presumably, 92\% of symptomatic infections are not reported to health authorities~\cite{Stanaway2016}. Low reporting rates can have many reasons, including low awareness levels and misdiagnosis~\cite{Bhatt2013,Simmons2012}.

Due to the rapid global spread of dengue as well as severe under-reporting, many countries are facing the threat of ongoing local transmission in the near future~\cite{Murray2013}. In non-endemic countries, local outbreaks are usually triggered by an imported case~\cite{Chang2015}, a person who acquired the disease overseas and transmitted the virus to local mosquitoes. To prevent ongoing dengue transmission in non-endemic countries, it is critical to forecast the importation of disease cases into these areas and  move from responsive containment of dengue outbreaks to proactive outbreak mitigation measures. 

The majority of existing models forecast relative rather than absolute risk of dengue importation and are unable to predict the total number of imported disease cases~\cite{Gardner2013,Semenza2014,Gardner2015}. The few models that can predict absolute numbers are region-specific rather than global~\cite{Wilder-Smith2014,Quam2015,Massad2018}. The most recently proposed model estimates the total number of imported dengue cases for 27 European countries~\cite{Massad2018}, however, the model has several limitations: (i)~Monthly incidence rates were based on dengue cases reported to the World Health Organisation (WHO) despite dengue being under-reported and the general consensus that the actual number of cases is much higher than the figures published by the WHO~\cite{Stanaway2016,Bhatt2013}; (ii)~Only 16 countries were considered as possible sources of importation. The authors reason that these 16 countries contribute 95\% of all global dengue cases, referring to numbers published by the WHO. Since African countries do not report to the WHO, and dengue remains an under-reported disease in many other countries~\cite{Standish2010,Kakkar2012,Vong2012,Wahyono2017}, it is likely that the percentage contribution to the number of global dengue cases by the 16 selected countries is strongly biased; (iii)~Seasonal distributions of dengue cases were inferred based on information from only two source countries (Latin American countries were assumed to have similar seasonalities to Brazil, while Thailand served as a proxy for countries in South-East Asia). The assertion that all countries within a given global region experience similar seasonal fluctuations in dengue infections is likely inaccurate. For example, dengue notifications peak between April and December in Thailand, while Indonesia reports the highest number of dengue cases from November to April~\cite{IAMAT}.

The contribution of this paper is twofold. First, we develop a network model that overcomes the limitations of previous models by employing global air passenger volumes, country-specific dengue incidence rates and country-specific temporal infection patterns. We construct weighted directed networks, using data collected by the International Air Transportation Association (IATA) to capture the movement of air passengers. We calculate monthly, country-specific dengue incidence rates by combining data from the Global Health Data Exchange~\cite{GlobalHealthDataExchange2017}, the most comprehensive health database, and known seasonal patterns in reported dengue infections~\cite{IAMAT}. Further, we distinguish between two categories of travellers: returning residents and visitors. The number of days people from these two categories  spend in an endemic country, and therefore the risk of being infectious on arrival, vary greatly. The model predicts the number of imported dengue cases per month for any given airport and can be applied with relative ease to other vector-borne diseases of global concern, such as malaria, Zika or chikungunya.

Second, we apply the model to infer time-varying, region-specific reporting rates, defined as the ratio of reported to actual infections. Dengue reporting rates vary greatly across space and time, often by several orders of magnitude, and hence are difficult to determine~\cite{Stanaway2016}. The usual approach towards estimating country-specific reporting rates is to carry out cohort or capture-recapture studies that can be costly, are time consuming and may be biased~\cite{Toan2015}. Consequently, dengue reporting-rates remain unknown for most countries~\cite{Stanaway2016}. 

In this paper we focus on those countries that are most at risk of dengue introduction, i.e. non-endemic countries with vector presence. These countries will have the greatest benefit from our model as knowledge about the likely arrival times and places of infected people is crucial to prevent local outbreaks.

\section*{Materials and methods}
CSIRO's human research ethics committee CSSHREC has approved this study (approval number: Ethics Clearance 142/16). All data were analysed anonymously and individuals cannot be identified.

\subsection*{IATA Data}
The International Air Transportation Association (IATA) has approximately 280 airline members who together contribute to approximately 83\% of all air traffic. Data is collected in form of travel routes, detailing the origin, destination and stopover airports. It contains over 10,000 airports in 227 different countries and dependencies.  For each route the total number of passengers per month is given. We do not have any information on stopover times and whether passengers are leaving the airport during their stopover and therefore assume that all passengers continue their journey to the final destination instantly. Table~\ref{tab:abbreviations} lists the IATA 3-Letter Codes used to abbreviate airports in the main manuscript. As the recorded itineraries do not include any travel on chartered flights, we compare the IATA passenger volumes to official airport passenger statistics~\cite{AirlineNetworkNewsandAnalysis2018,CivilAviationAdministrationofChina2016,MinistryofLandInfrastructureTransportandTourism,MinistryofTransportationRepublicofIndonesia2018,AirportsCouncilInternational2015,DirectorateGeneralofCivilAviation-Kuwait2015,EANA-NavegacionAereaArgentina2016,AngkasaPuraII2015,CivilAviationAuthorityofthePhilippines2015,InstitutodeEstadisticasdePuertoRico2015,MalaysiaAirports2015,Zambia:TransportDataPortal2016,CaliforniaDepartmentofTransportation2016,AirportsAuthoriyofIndia2016,AssociazioneItalianaGestoriAeroporti2016,OfficeNationaldesAeroports2017,AirportsCompanySouthAfrica2018,EgyptianHoldingCompanyforAirportsandAirNavigation2015,AirportsCouncilInternational2018} to quantify the potential discrepancies between actual travel patterns and that reported by IATA. Table~\ref{tab:excluded_countries} lists the countries where the difference in passenger numbers is greater than 15\% (at country level) and countries where airport statistics were not available and the tourist data suggests inaccuracies in the IATA data (i.e. the number of tourists arriving in a particular country is larger than the total number of passengers arriving). We also excluded Singapore as a source of importation for Australia for the following reason: The Department of Home Affairs publishes Arrival Card data~\cite{DepartmentofHomeAffairs2017} that can be used to validate the IATA data. A comparison of the monthly travel volume from Singapore to Australia revealed that the IATA data overestimates travel volumes by approximately 112\% on average in 2011 and 2015. This may be due to individuals who travel from other countries to Singapore and then directly continue to Australia and do not book their entire trip in one itinerary (this would be recorded as two separate trips in the IATA data that cannot be linked to each other). Due to this large discrepancy in the travel data we believe that our model will significantly overestimate the number of dengue infections imported from Singapore, and therefore exclude it as a source country for Australia.

\subsection*{The air transportation network}
We begin by constructing twelve weighted, directed networks, using IATA data, to represent the monthly movement of air passengers during a given year. The networks are denoted $\mathcal{G}_m=(V,E)$, with $m=1,\dots,12$ indicating the month of the year. The node set $V$ comprises more than 10,000 airports recorded by IATA. To distinguish the travellers by their country of embarkation, we represent the edges of the network as ordered triples, $(i,j,\omega_{i,j}(c,k))\in E$, where $i,j\in V$ and $\omega_{i,j}(c,k)$ is a function that outputs the number of passengers who initially embarked in country $c$ with final destination airport $k$ and travel from airport $i$ to airport $j$ as part of their journey.

\subsection*{Incidence rates and seasonal distributions}
Calculating the number of infected passengers requires daily infection probabilities. We derive these from country-level yearly estimates of symptomatic dengue incidence rates that are published together with their 95\% confidence intervals by the Global Health Data Exchange~\cite{GlobalHealthDataExchange2017}. The estimates are obtained using the model published in~\cite{Stanaway2016} and account for under-reporting. 

We first deduce monthly incidence rates using information on dengue seasonality published by the International Association for Medical Assistance to Travellers~\cite{IAMAT}. To do so we associate a weight with each month that indicates the intensity of transmission. To assign the weights we use a modified cosine function with altered period that matches the length of the peak-transmission season. The function is shifted and its amplitude adjusted so that its maximum occurs midway through the peak-season with value equal to the length of the peak-season divided by $2\pi$. The months outside the peak-season receive a weight of one if dengue transmission occurs year around and a weight of zero if dengue transmission ceases outside the peak-season. The weights are then normalised and multiplied by the yearly incidence rate for the corresponding country. Normalising the weights ensures that the sum of the monthly incidence rates is equal to the yearly incidence rate. To calculate the lower and upper bounds of the monthly incidence rates, we multiply the normalised weights by the lower and upper bounds of the 95\% confidence interval given for the yearly incidence rates. 

The average probability, $\beta_{c,m}$, of a person becoming infected on any given day during month $m$ in country $c$ is then given by 

\begin{equation}\label{eqn:ratetoprob}
\beta_{c,m} = 1 - e^{-\gamma_{c,m}/d_m},
\end{equation}

\noindent
where $\gamma_{c,m}$ is the monthly dengue incidence rate in country $c$ during month $m$ and $d_m$ is number of days in month $m$. Note that Equation~\eqref{eqn:ratetoprob} converts the daily incidence rate into the probability of a single person becoming infected with dengue on any given day during month $m$. 

\subsection*{Inferring the number of infected passengers}
Next, we present a mathematical model that approximates the number of dengue-infected people for each edge in the network $\mathcal{G}_m(V,E)$. The time between being bitten by an infectious mosquito and the onset of symptoms is called the intrinsic incubation period (IIP). This period closely aligns with the latent period, after which dengue can be transmitted to mosquitoes~\cite{Chan2012}. The IIP lasts between 3 and 14 days (on average 5.5 days) and was shown to follow a gamma distribution of shape 53.8 and scale equal to 0.1~\cite{Chowell2007}. After completion of the IIP a person is infectious for approximately 2 to 10 days (on average 5 days)~\cite{Gubler1995,Chowell2007}. The length of the infectious period was shown to follow a gamma distribution of shape 25 and scale equal to 0.2~\cite{Chowell2007}. We denote the sum of the IIP and the infectious period by $n$, which is rounded to the nearest integer after the summation. For travellers to import the infection from country $c$ into a new location $r$ they must have been infected with dengue within the last $n-1$ days of their stay in country $c$. We now consider the following two cases: $t_{c}\geq n-1$ and $t_{c} < n-1$, where $t_{c}$ is number of days spent in country $c$ before arriving in region $r$. Since we do not know the exact date of arrival for travellers, we assume that arrival and departure dates fall within the same month and hence $\beta_{c,m}$ is the same for every day during the travel period.

If $t_{c} \geq n-1$, that is the individual spent more time in country $c$ than the sum of the lengths of the IIP and the infectious period, the probability of not being infected on return is equal to $(1-\beta_{c,m})^{t_{c}}+\left[1-(1-\beta_{c,m})^{t_{c}-(n-1)}\right]$. The first term covers the possibility that the individual did not get infected whilst staying in country $c$ and the second term covers the possibility that the individual got infected and recovered before arriving at a given airport (see Fig~\ref{fig:diagram}). Hence, the probability of a person, who arrives at a given airport from country $c$ during month $m$, being infected with dengue is given by

\begin{eqnarray}
p_{c,m} &=& 1-\left[(1-\beta_{c,m})^{t_{c}}+1-(1-\beta_{c,m})^{t_{c}-(n-1)}\right]\nonumber\\
&=& (1-\beta_{c,m})^{t_{c}-(n-1)}-(1-\beta_{c,m})^{t_{c}}.
\end{eqnarray}

If $t_{c} < n-1$, that is the individual spent less time in country $c$ than the sum of the lengths of the IIP and the infectious period, the probability of not being infected on return is equal to $(1-\beta_{c,m})^{t_{c}}$, which covers the possibility that the individual did not get infected whilst staying in country $c$. Since $t_{c} < n-1$, the probability of recovery before arriving at a given airport is zero. Hence, the probability of a person, who arrives from country $c$ at a given airport during month $m$, being infected with dengue is given by

\begin{equation}
p_{c,m} = 1-(1-\beta_{c,m})^{t_{c}}.
\end{equation}

We distinguish between two different types of travellers arriving at a given airport of region $r$: returning residents and visitors. We define a returning resident as a traveller who resides in region $r$ and a visitor as a traveller who resides in country $c$ and visits region $r$. Returning residents are expected to have stayed a couple of weeks in the endemic country, while visitors may have spent their whole life in the country.

Since we lack information on how long each individual spent in country $c$ before arriving at an airport of region $r$, we substitute parameter $t_{c}$ by $\langle t\rangle_{c}^{res}$ if the person is a returning resident, $\langle t\rangle_{c}^{res}$ being the average number of days a returning resident spends in country $c$ before returning home. If the person is a  visitor, parameter $t_{c}$ is substituted by $\langle t\rangle_{c}^{\textrm{vis}}$, the average number of days a visitor spends in country $c$ before arriving at an airport of region $r$. We distinguish between returning residents and visitors since $\langle t\rangle_{c}^{\textrm{res}}\ll \langle t\rangle_{c}^{\textrm{vis}}$. 

We assume that the length of stay for returning residents follows a normal distribution with mean equal to 15 days and standard deviation of 2, i.e. $\langle t\rangle_{c}^{res} \sim \mathcal{N}(15, 2)$. A previous study has shown that employees around the world are on average entitled to approximately 15 days of annual leave~\cite{Messenger2007}. On the other hand, visitors likely spent all their lives in the endemic country. We assume that $\langle t\rangle_{c}^{vis} \sim \mathcal{N}(\mu_{vis},0.1\mu_{vis})$, where $\mu_{vis}$ is equal to $c$'s median population age. Median population ages by country are published in the World Factbook by the Central Intelligence Agency~\cite{CentralIntelligenceAgency2017}.

For simplicity we do not take immunity to the different dengue strains into consideration.

\subsection*{Proportion of returning residents and visitors}
Lastly, we need to infer the proportions of returning residents and visitors. As this information is not contained in the IATA itineraries, we use international tourism arrival data from the World Tourism Organisation~\cite{WorldTourismOrganisation2018}. The data contains the yearly number of international tourist arrivals by air for each destination country. From the IATA data we calculate the total number of arrivals per year for each country and hence can infer the ratio of visitors to returning residents. As we lack sufficient data, we assume that the ratio of visitors to residents is the same for each month.

\subsection*{Calculating the absolute number of infected passengers}
Given the above, we can now determine the number of infected passengers $I_{k,m}$ arriving at airport $k$ during month $m$ as follows:

\begin{equation}
I_{k,m} = \sum_{i,j,c}\omega_{i,j}(c,k)\left[qp_{c,m}^{res} + (1-q)p_{c,m}^{vis}\right],
\end{equation}

\noindent
where $q$ is the proportion of residents inferred from the international tourism arrival data, 

\begin{strip}
\begin{equation}
p_{c,m}^{res}= \begin{cases}
(1-\beta_{c,m})^{\langle t\rangle_{c}^{res}-(n-1)}-(1-\beta_{c,m})^{\langle t\rangle_{c}^{res}} & \langle t\rangle_{c}^{res}\geq n-1\\
&\\
1-(1-\beta_{c,m})^{\langle t\rangle_{c}^{res}} & \langle t\rangle_{c}^{res}< n-1,
\end{cases}
\end{equation}

\noindent
and

\begin{equation}
p_{c,m}^{vis}= \begin{cases}
(1-\beta_{c,m})^{\langle t\rangle_{c}^{vis}-(n-1)}-(1-\beta_{c,m})^{\langle t\rangle_{c}^{vis}} & \langle t\rangle_{c}^{vis}\geq n-1\\
&\\
1-(1-\beta_{c,m})^{\langle t\rangle_{c}^{vis}} & \langle t\rangle_{c}^{vis}< n-1.
\end{cases}
\end{equation}
\end{strip}

\subsection*{Evaluation of the model’s uncertainty}
We performed a thousand runs of the model for each edge in the network, drawing the parameters from their respective distributions, to calculate the mean and standard deviation of dengue-infected passengers. In addition, we have conducted a global sensitivity analysis to identify the model parameters with the greatest influence. We used Sobol's method~\cite{Sobol2001} with 100,000 samples to carry out the sensitivity analysis. The parameter ranges are shown in Table~\ref{tab:parameters}. The analysis was done with SALib~\cite{Herman2017}, an open-source Python library.

\begin{table}[!ht]
		\centering
		\begin{tabular}{ll}
			\toprule
			Parameter & Range\\ 
			\midrule
			$\beta_{c,m}$&[0.000001, 0.000445]\\ 
			$t_c$ (days)&[1, 29200]\\ 
			$n$ (days)&[5, 24]\\ 
			\bottomrule
		\end{tabular}
		\caption{\bf The model parameter ranges used in Sobol's method.}\label{tab:parameters}
\end{table}

% Results and Discussion can be combined.
\begin{figure*}[!h]
	\centering
	\includegraphics[width=0.8\linewidth]{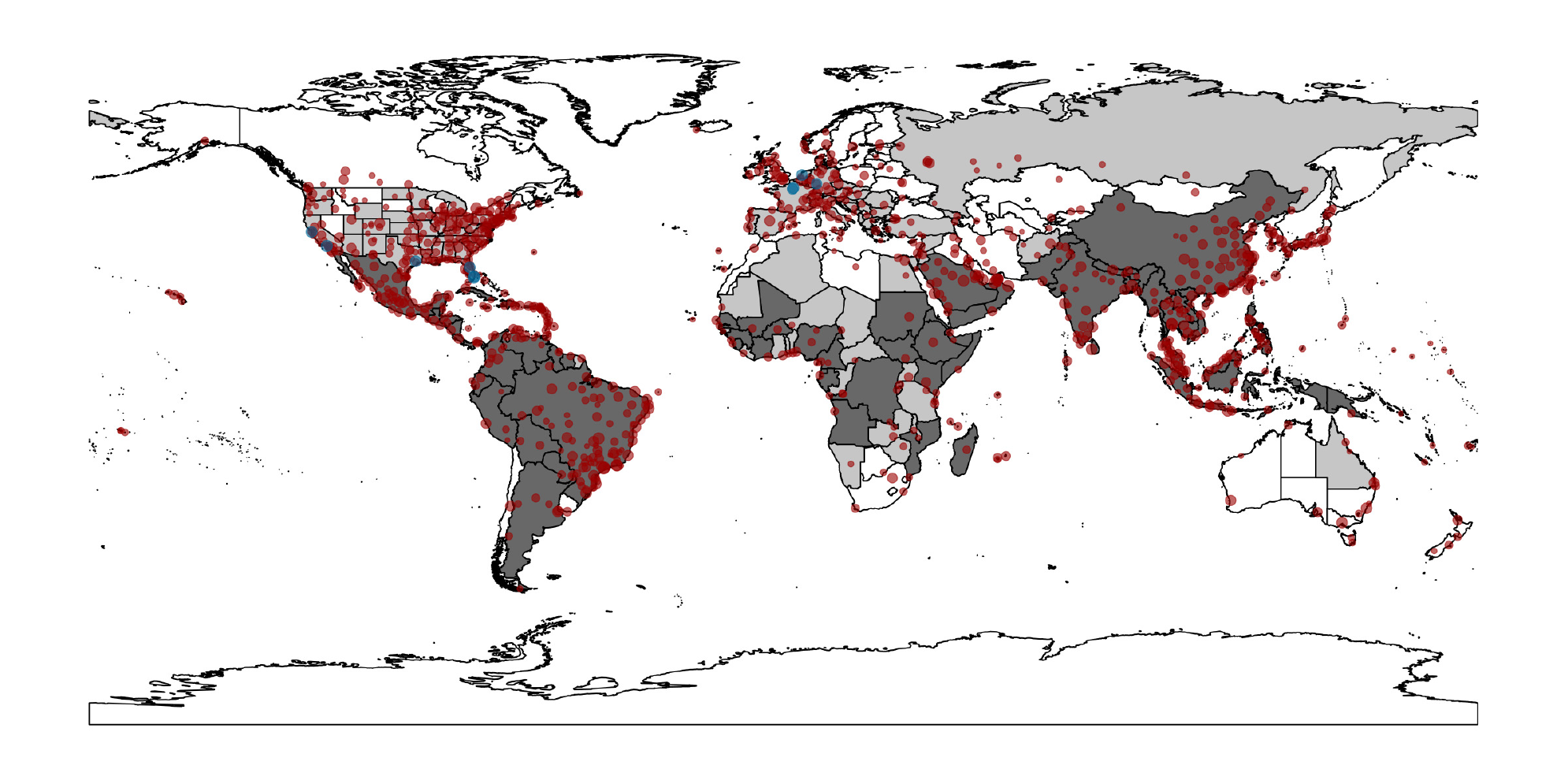}
	\caption{{\bf Predicted dengue importations for August 2015.} \normalfont The map shows the output of our model for August 2015.The area of a node increases with the number of dengue cases imported through the corresponding airport. Airports that are predicted to not receive any infections are not shown on the map. Endemic countries are coloured dark grey. Countries that are non-endemic and where dengue vectors \emph{Aedes aegypti} and/or \emph{Aedes albopictus} are present are coloured in light grey. The blue circles correspond to the top ten airports identified in Fig~\ref{fig:top10}. The map was created with the Python GeoPandas package and publicly available shapefiles from Natural Earth (\url{http://www.naturalearthdata.com/}).}\label{fig:map}
\end{figure*}

\section*{Results}
We run our model for two different years to explore the robustness of the proposed methodology. Specifically, the analysis is conducted for 2011 and 2015. The results for the year 2015 are presented in the main manuscript, while the results for 2011 are presented in the supplementary material. Fig~\ref{fig:map} shows the number of predicted imported dengue infections per airport for August 2015, where the area of a node increases with the number of dengue cases imported through the corresponding airport. The map clearly shows that many non-endemic regions where the dengue-transmitting vectors \emph{Aedes aegypti} or \emph{Aedes albopictus} are present (coloured in light grey) have airports that are predicted to receive a high number of dengue infections. For a list of dengue endemic and non-endemic countries see Table~\ref{tab:endemic_vector}. As resources for the control and prevention of dengue are often limited~\cite{Morrison2008}, these countries face a high risk of future endemicity.

\begin{figure}[!h]
	\centering
	\includegraphics[width=0.9\linewidth]{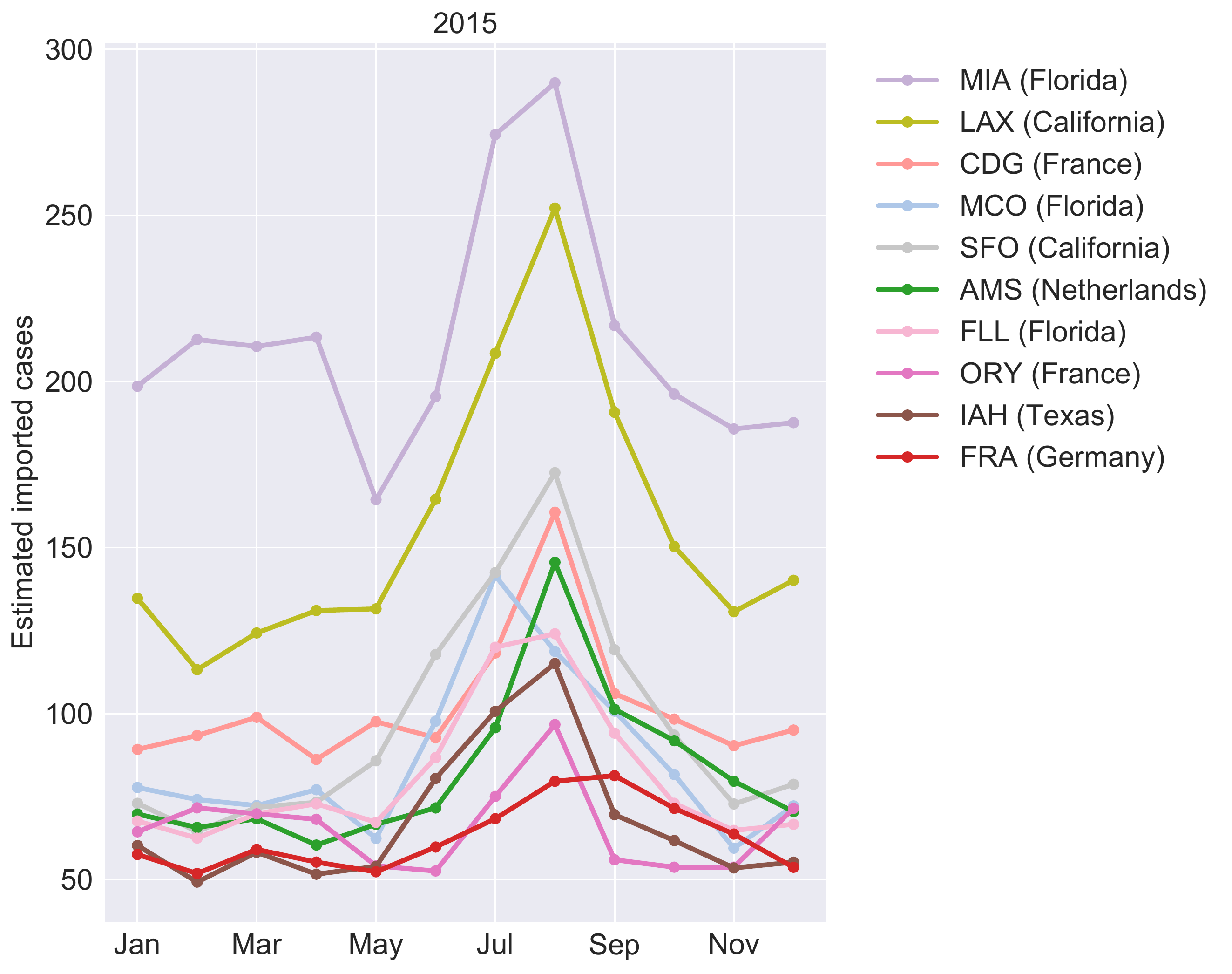}
	\caption{{\bf Predicted monthly dengue importations by airport for 2015.} \normalfont The number of predicted imported dengue infections for the top ten airports in non-endemic countries/states with vector presence for each month in 2015. A break in a line indicates that the corresponding airport was not amongst the top ten during the respective month. Airports are abbreviated using the corresponding IATA code. A full list of abbreviations can be found in the supplementary material (see Table~\ref{tab:abbreviations}).}
	\label{fig:top10}
\end{figure}

In Fig~\ref{fig:top10} and Fig~\ref{fig:Top10_2011} we plot the number of predicted dengue importations over time for the ten airports that receive the highest number of cases, lie in non-endemic regions with vector presence and where local cases have been reported in the past (more detailed plots with confidence intervals are shown in Fig`\ref{fig:Top10_CI}). While the majority of airports listed in Fig~\ref{fig:top10} and Fig~\ref{fig:Top10_2011} are predicted to receive between 50 and 150 cases each month, Miami International Airport (MIA) is estimated to receive between 146 and 309 cases each month during both years. With Orlando International Airport (MCO) and Fort Lauderdale–Hollywood International Airport (FLL) also represented amongst the airports with the highest number of imported cases, Florida faces a high risk of local dengue outbreaks. Los Angeles International Airport (LAX) is predicted to receive the second highest number of imported cases. In 2011 its monthly predictions vary between 97 and 205 cases and in 2015 between 113 and 253 cases. The remaining airports listed in Fig~\ref{fig:top10} and Fig~\ref{fig:Top10_2011} are located in France, Germany, the Netherlands, Texas, and Queensland, Australia. A full ranking of all airports located in non-endemic countries with vector presence can be found in Table~\ref{tab:annual_imported} of the supplementary material.

\begin{figure*}[!h]
	\centering
	\includegraphics[width=0.9\linewidth]{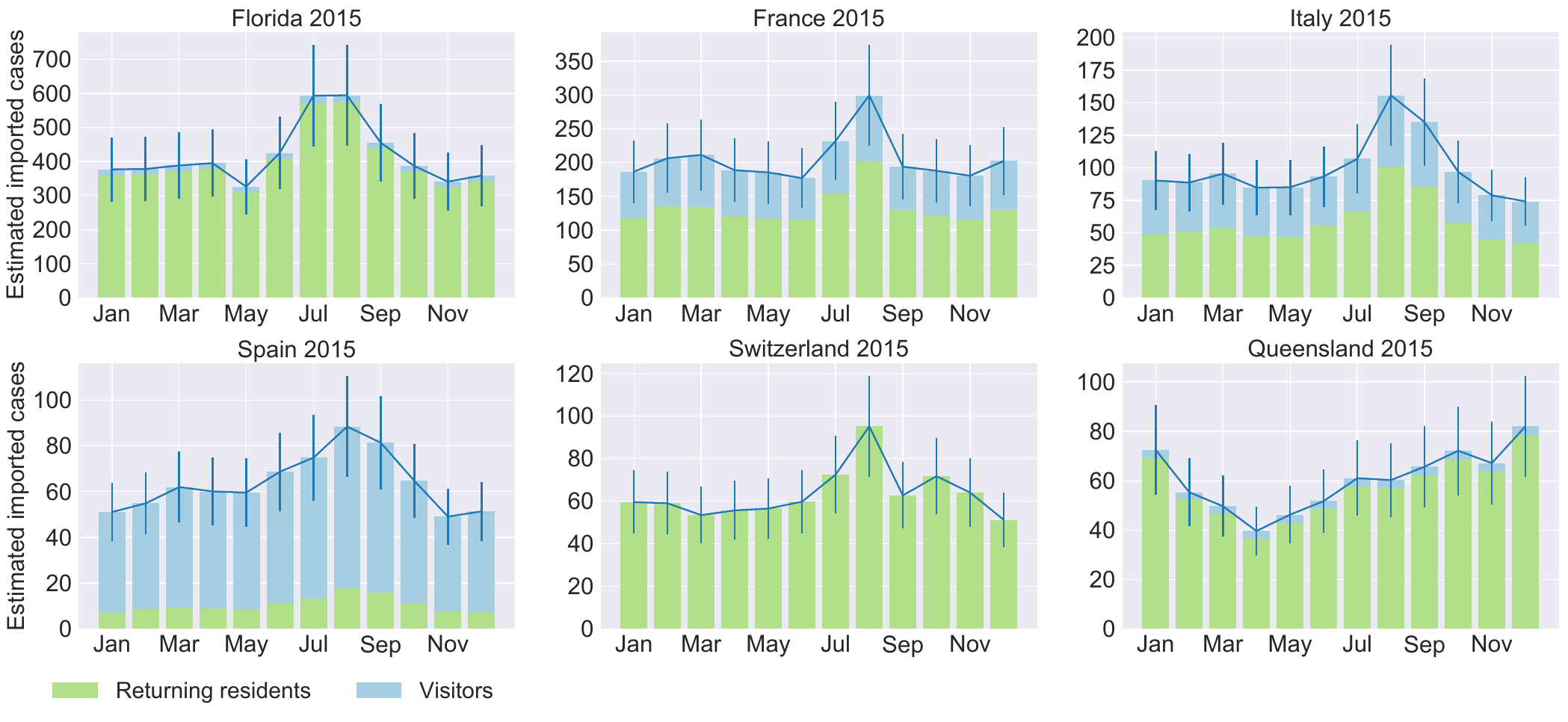}
	\caption{{\bf Predicted dengue infections imported by returning residents and visitors in 2015.} \normalfont Here we show the results for non-endemic countries/states with vector presence with the highest number of predicted imported dengue cases in 2015. The bars are stacked to distinguish between returning residents (green) and visitors (blue). The blue solid line corresponds to the total number of imported cases. The error bars correspond to the model's coefficient of variation (see Material and methods). The six countries were selected because they are predicted to receive the highest number of dengue importations, are non-endemic and dengue vectors are established.}
	\label{fig:visRes}
\end{figure*}

In addition to calculating the number of imported dengue infections per airport, the model further provides the number of infected passengers travelling between any two airports, thus revealing common importation routes. Table~\ref{tab:routes} and Table~\ref{tab:routes2011} list the routes that carry the highest number of infected passengers whose final destinations lie in non-endemic countries with vector presence. Table~\ref{tab:routesS} lists the routes that carry the highest number of infected passengers whose final destinations lie in non-endemic countries irrespective of whether vectors are present. For example, the route between Denpasar and Perth is ranked third in 2011 in Table~\ref{tab:routesS}, but it is not considered in the ranking shown in Table~\ref{tab:routes2011}, as there are no vectors in Perth. Fig~\ref{fig:importation_routes_Aug_2015} shows a map of all importation routes into non-endemic countries with vector presence.

\begin{table}[!ht]
		\centering
		\begin{tabular}{llll}
			\toprule
			Orig. & Dest. & Pax & Month\\
			\midrule
			SJU (Puerto Rico)& MCO (Florida) & 51 & Jul\\ 
			PTP (Guadeloupe) & ORY (France) & 37 & Aug \\ 
			FDF (Martinique)& ORY (France)& 34 & Aug\\ 
			SJU (Puerto Rico)& FLL (Florida) & 32 & Jul\\ 
			TPE (Taiwan)& LAX (California) &31 &Aug\\ 
			GRU (Brazil)& MIA (Florida) & 29 & Apr\\ 
			DEL (India)& KBL (Afghanistan) & 27 & Aug\\ 
			GDL (Mexico)& LAX (California)& 24 & Aug\\ 
			CUN (Mexico)& MIA (Florida) & 24 & Aug\\ 
			CUN (Mexico)& LAX (California)& 22 & Aug\\ 
			\bottomrule
		\end{tabular}
		\caption{{\bf The ten routes with the highest predicted number of dengue-infected passengers with final destinations in non-endemic countries with vector presence.} The table lists the direct routes with the highest predicted volume of dengue-infected passengers who continue to travel to non-endemic regions with vector presence and where local outbreaks have been reported in the past. The last column records the month during which the highest number of infected passengers are predicted.}\label{tab:routes}
\end{table}

In both years the highest predicted number of infected passengers are recorded during the northern hemisphere's summer. The route between S\~{a}o Paulo International Airport (GRU) and Miami International Airport (MIA) is the exception, where the highest number of infected passengers is predicted during April. The routes with the highest estimated number of dengue-infected passengers terminate at airports in countries that are non-endemic and where dengue-transmitting vectors are present.

\subsection*{Returning residents and visitors}
Next, we aggregate airports by country/state to predict the number of imported dengue infections on a coarser level. For non-endemic countries that cover an area larger than 5,000,000 km$^2$ and where dengue vectors are present we aggregate airports by state. These countries are Russia, the United States of America and Australia. The comparison between passenger volumes recorded by IATA and official airport statistics indicated that the IATA data for Russia may be inaccurate, i.e. the difference in passenger numbers is larger than 15\% (see Material and Methods). Hence, we did not perform a state-level analysis for this country. In Australia vectors are present only in Queensland~\cite{Beebe2009}. While vectors have been observed in more than 40 different US states, autochthonous cases have been reported only in California, Florida, Hawaii and Texas~\cite{Hahn2016}.

\begin{figure*}[!h]
	\centering
	\includegraphics[width=0.9\linewidth]{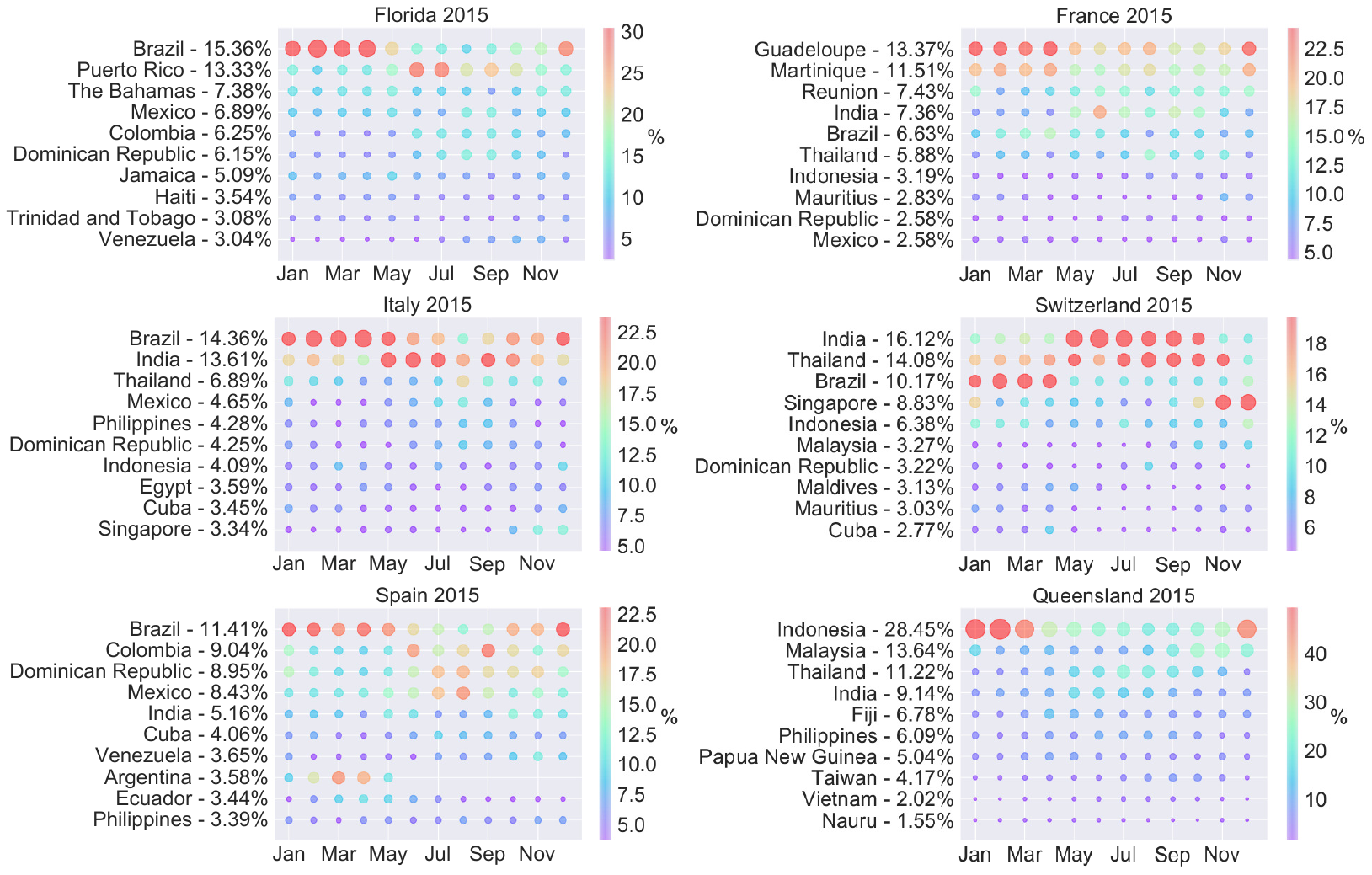}
	\caption{{\bf Predicted percentage contribution of dengue importations by country of acquisition in 2015.} The predicted percentage contribution by source country and month in 2015. The size and colour of the circles indicate the percentage contribution of the corresponding country to the total number of imported cases. The $y$-labels indicate the yearly percentage contribution of the corresponding source country.}
	\label{fig:acquisition}
\end{figure*}

Our model separately calculates the number of dengue-infected people amongst returning residents and visitors and hence we can identify which of these groups is more likely to import the disease into a given country or state. Fig~\ref{fig:visRes} and Fig~\ref{fig:visitorsResidents2011} show the results for six non-endemic countries/states with vector presence that are predicted to receive the highest number of dengue importations each month. Results for the remaining countries and states are shown in Figs~\ref{fig:US2011} -~\ref{fig:Africa}. We observe that the contributions of returning residents and visitors to the total number of imported dengue infections is predicted to vary greatly between the different countries and states. In Florida and Queensland returning residents are predicted to be the main source of dengue importation. In France and Italy approximately one third of all dengue infections are predicted to be imported by visitors while in Spain visitors import around 75\% of all imported cases. For Switzerland we do not have any information about the ratio of returning residents to visitors. For the United States there is evidence in the form of surveillance reports that returning residents are indeed the main contributors to dengue importations~\cite{VanDodewaard2015}. For Queensland we predict that 95\% and 94\% of infections were imported by returning residents in 2011 and 2015, respectively. Our predictions are supported by Queensland's dengue notification data (provided by Queensland Health), showing that 97\% and 92\% of all dengue importations in 2011 and 2015, respectively, were imported by returning residents.

\subsection*{Countries of acquisition}
In addition to being able to distinguish between returning residents and visitors, the model also divides the imported cases according to their places of acquisition. Fig~\ref{fig:acquisition} and Fig~\ref{fig:percentage_acquisition_2011} show the model's estimated percentage contribution of dengue importations by source country. 

\begin{figure*}[!h]
	\centering
	\includegraphics[width=0.9\linewidth]{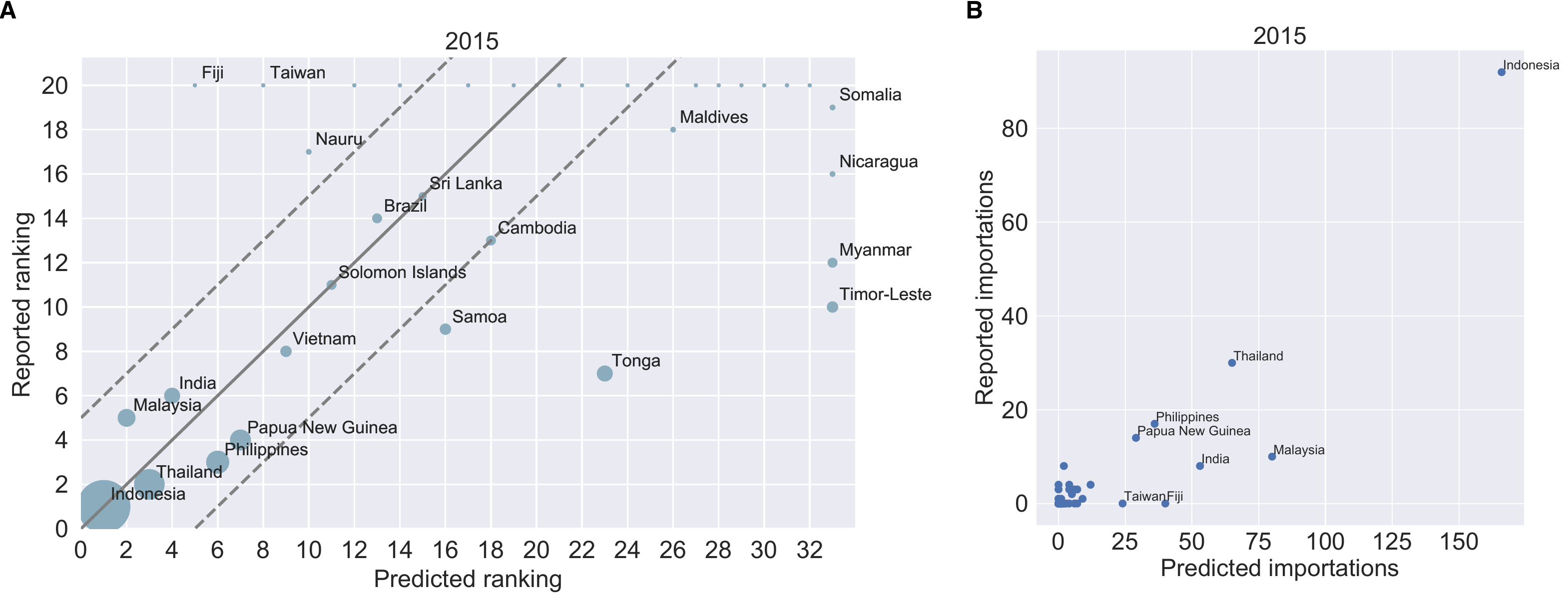}
	\caption{{\bf Rank-based validation and correlation between reported and predicted imported cases for Queensland in 2015.} ({\bf A}) Countries are ranked by the total number of predicted and reported imported dengue cases. The reported ranking is then plotted against the predicted ranking. Countries that were ranked by the model, but did not appear in the dataset receive a rank of $i+1$, were $i$ is the number of unique importation sources according to the dengue case data. Similarly, countries that appeared in the data and were not ranked by the model receive a rank of $i+1$. For circles that lie on the $x=y$ line (grey solid line) the predicted and reported rankings are equal. Circles that lie between the two dashed lines correspond to countries with a difference in ranking that is less than or equal to five. The circle areas are scaled proportionally to the number of reported cases that were imported from the corresponding country. Spearman's rank correlation coefficient between the absolute numbers of reported and predicted importations is equal to 0.6. ({\bf B}) The absolute number of reported dengue importations are plotted against the absolute number of predicted importations.}
	\label{fig:rankValidation}
\end{figure*}

Florida is predicted to import most infections from the Caribbean and Latin America, with infections acquired in Puerto Rico (PRI) predicted to peak during June and July and infections acquired in Brazil predicted to peak between January and April. We hypothesise that Florida receives such a high number of imported dengue cases due to its close proximity to the Caribbean, which has been endemic since the 1970s~\cite{Amarakoon2008}. France is predicted to receive many infections from the Caribbean, in particular from Martinique and Guadeloupe which are French overseas regions and hence a high volume of air traffic from these regions to metropolitan France is expected. These predictions align with the fact that outbreaks of dengue in France coincide with outbreaks in the French West Indies, where most reported cases are acquired~\cite{Vasquez2018,LaRuche2013}. In Italy the model predicts that the most common countries of acquisition are India and Brazil. India and Brazil are also the most common countries of acquisition for Switzerland in 2011. In 2015 Switzerland is predicted to receive most of their dengue importations from India and Thailand. Spain is predicted to import the majority of infections from Latin America and the Caribbean. For Queensland the model predicts that imported cases are acquired mostly in South-East Asia with Indonesia being the largest source. This is in agreement with previous studies~\cite{Warrilow2012} and the dengue case data that was provided by Queensland Health. In addition, we performed a rank-based validation of these results.

We obtained dengue case data from Queensland Health, which records the places of acquisition for each dengue case reported in Queensland. We rank the countries of acquisition by the total number of predicted and reported dengue-infected people who arrive in Queensland. We then plot the reported ranking against the predicted ranking. In addition, we plot the absolute number of reported importations against the absolute number of predicted importations and calculate Spearman's rank correlation coefficient. Fig~\ref{fig:rankValidation} and Fig~\ref{fig:rankingQLD2011} show the results.

The rank-based validation of our model demonstrates that overall, the model captures the different importation sources well. It does particularly well for the countries from which Queensland receives the most infections. Spearman's rank correlation coefficient is equal to 0.6 for the year 2015 and equal to 0.58 for the year 2011. Below we explain some of the differences between the data and the model output. 

For the rank-based validation the two largest outliers in both years are Fiji and Taiwan. The predicted ranking for Fiji in 2011 is 2, while the reported ranking is 10. In 2015 we estimate Fiji to be ranked fifth, however no cases were reported in 2015 and hence Fiji is ranked last amongst the reported cases. According to the Fijian government tourists are less likely to contract the disease than local residents as they tend to stay in areas that are not infested by \emph{Aedes aegypti} mosquitoes~\cite{TheFijianGovernment2014} or where there is likely considerable control effort undertaken by tourism accommodation operators. Since the incidence rates incorporated into our model do not distinguish between different regions of a source country, the model is unable to account for such nuances. In 2011 and 2015 we estimate Taiwan to be ranked seventh and eighth, respectively, however no cases were reported in both years. This result is surprising as dengue occurs year-round in Taiwan~\cite{IAMAT} and approximately 44,000 and 16,000 Queensland residents travelled to Taiwan in 2011 and 2015, respectively.

\begin{table*}[!h]
	\centering
	\begin{tabular}{lccccc}
		\toprule
		& Dec-Feb & Mar-May & Jun-Aug & Sep-Nov & Yearly\\ 
		\midrule
		Queensland & 32.4&48.9&18.6&22.6&28.6\\ 
		Spain&   14&14&31.7&26.3&23.5\\ 
		Italy  &  4.5&6.8&9.2&13.1&9\\ 
		France  &  3.8&6.9&9.7&7.1&7.2\\ 
		Florida   &  0.9&0.7&1.2&2.7&1.4\\
		\bottomrule
	\end{tabular}
	\caption{{\bf Yearly and seasonal reporting rates of imported cases in 2015.} The table shows the estimated reporting rates of imported cases for Queensland, Spain, Italy, France and Florida. We estimate the reporting rates by using a least squares linear regression without intercept.}\label{tab:reportingRates}
\end{table*}

Some of the differences between the observed percentages and the predicted percentages can be explained by under-reporting. It is possible that dengue awareness among travellers to one country is greater than the awareness amongst travellers to another country. Travellers with higher awareness levels are more likely to report to a doctor if feeling unwell after their return.

\subsection*{Country-specific reporting rates}
The reporting rate of a disease is defined as the ratio of reported infections to actual infections. Dengue reporting rates vary greatly across space and time and are difficult to determine~\cite{Stanaway2016}. The usual approach to estimating country-specific reporting rates is to carry out cohort or capture-recapture studies that can be costly, are time consuming and may be biased~\cite{Toan2015}. 

We utilised our model to infer country- and state-specific reporting rates of imported cases by performing a least squares linear regression without intercept.

Table~\ref{tab:reportingRates} and Table~\ref{tab:reportingRates2011} show the estimated yearly and seasonal reporting rates of imported cases for Queensland, Florida, France, Italy and Spain. To distinguish locally acquired and imported cases in Queensland, we use case-based data from Queensland Health where the country of acquisition is recorded. Travel-related dengue cases reported in Europe are published by the European Centre for Disease Prevention and Control (\url{http://ghdx.healthdata.org/gbd-results-tool}). Data for Florida is available from the Florida Department of Health (\url{http://www.floridahealth.gov/diseases-and-conditions/mosquito-borne-diseases/surveillance.html}). 

The results show that estimated reporting rates of imported cases are highest in Queensland, in particular during autumn. This is expected as dengue awareness campaigns are intensified between November and April~\cite{QueenslandHealth2015}. In contrast, Florida has the lowest dengue reporting rate (1.3\% in 2011 and 1.4\% in 2015). This finding is supported by a previous study which found that awareness levels in Florida are extremely low~\cite{Hayden2015}. The estimated reporting rates for the European countries are also low; however, the model predicts a substantial increase from 2011 to 2015. The question why reporting rates in Queensland are higher is challenging to answer, as we do not have any information about the true number of imported cases. However, Queensland has one of the best dengue prevention programs in the world. According to Queensland Health, other states and countries frequently ask for training and advice regarding surveillance and awareness campaigns.

\subsection*{Model uncertainty}
We found that the average coefficient of variation of our importation model is 19.5\% across both years. That is, the model's standard deviation is on average equal to 19.5\% of its mean. Fig~\ref{fig:cv} shows the distribution of the coefficient of variation for several destinations. 

The results from the global sensitivity analysis show that $t_c$ is the most important of the three model parameters with a total-order index of 0.94 (see Fig~\ref{fig:sensitivity}). The different values of the first-order and total-order indices indicate interaction between the model parameters. The second-order indices show that there is significant interaction between parameters $t_c$ and $\beta_{c,m}$ with a second-order index of 0.19, as well as between parameters $t_c$ and $n$ with a second-order index of 0.1.

Since the range of parameter $t_c$ is large ([1, 29200] days), we performed the sensitivity analysis again for a shorter range of values ([1, 30] days) that is more realistic for returning residents who spend their holidays in an endemic country. In this case, parameter $\beta_{c,m}$, with a total-order index of 0.6, is more important than $t_c$, which has a total-order index of 0.35 (see Fig~\ref{fig:sensitivity}). The second-order indices show that there is still significant interaction between parameters $t_c$ and $\beta_{c,m}$ with a second-order index of 0.06, and between parameters $t_c$ and $n$ with a second-order index of 0.07.

\section*{Discussion}
To mitigate the risk of outbreaks from importation of dengue into non-endemic regions it is critical to predict the arrival time and location of infected individuals. We modelled the number of dengue infections arriving each month at any given airport, which enabled us to estimate the number of infections that are imported into different countries and states each month. In addition, the model determines the countries of acquisition and hence is able to uncover the routes along which dengue is most likely imported. Our results can also be used to estimate country- and state-specific reporting rates of imported cases.

Such knowledge can inform surveillance, education and risk mitigation campaigns to better target travellers along high risk importation routes at the most appropriate times. It will also help authorities to more efficiently surveil those airports with the highest risk of receiving dengue-infected passengers.

The model proposed here overcomes many of the shortcomings of previous models, however, it is not without limitations. Validation through comparison of reported cases to predicted cases is infeasible due to the high degree of under-reporting. However, we demonstrate that the coefficient of variation of the model with 19.5\% on average is low (see Material and Methods). A rank-based validation for Queensland confirmed that the different importation sources are accurately predicted. 

Incidence rates may vary considerably from region to region within the same country~\cite{TheFijianGovernment2014} and higher resolution data could improve the model's predictions, as it would better reflect the export of dengue cases from the individual regions. Region-specific incidence rates can, for instance, be combined with spatial patterns of the visiting frequency of travellers to determine the likelihood of travellers to export dengue out of endemic countries. Additional data on individuals' travel behaviour may also be beneficial, as it can be analysed to improve the estimation of the average time that a person has spent in a specific country before arriving at a given airport. Our assumption that returning residents and visitors are exposed to the same daily incidence rates is a simplification. Further details on the types of accommodation, for example, resorts vs local housing, could also be used to inform the daily incidence rates, due to variations in vector control. The global sensitivity analysis has revealed that $t_c$, the number of days a traveller has spent in country $c$, is the most important model parameter. Hence, additional data on individuals' travel behaviour may substantially improve the model. Knowledge about the exact age of visitors who reside in non-endemic countries would  also improve the model. Currently, we assume that the age of a visitor is equal to the median age of the population of the country in which the visitor resides. In reality, the age of air passengers may differ from the median age, especially for developing countries.

In temperate regions local conditions may not allow for dengue to be transmitted during the winter months. Thus, even a large number of imported cases during those months would not trigger local outbreaks. Variable seasonality patterns due to El Ni\~{n}o Southern Oscillation can affect the spread of dengue in tropical and subtropical regions. An interesting direction for future research is to combine the here proposed model with knowledge of local conditions and weather phenomena like El Ni\~{n}o Southern Oscillation to evaluate the risk of local outbreaks. In this work we studied dengue importation via air travel. In future, we will also consider other modes of transportation to develop a more comprehensive model.

\section*{Data availability statement}
The air travel data used in this study are owned by a third party and were licensed for use under contract by International Air Travel Association (IATA)- Passenger
Intelligence Services (PaxIS): \url{http://www.iata.org/services/statistics/intelligence/paxis/Pages/index.aspx}. The same data can be purchased for use by any other researcher by contacting: Phil GENNAOUI Regional Manager - Aviation Solutions (Asia Pacific) Tel: +65 6499 2314 | Mob: +65 9827 0414 gennaouip@iata.org | www.iata.org

Monthly dengue incidence rates for all countries for the years 2011 and 2015 are available as supplementary information files.

Median population age data is publicly available from the Central Intelligence Agency at \url{https://www.cia.gov/library/publications/resources/the-world-factbook/fields/343.html}.

International tourism arrival data used in this study are owned by a third party and cannot be shared publicly. The data is available for purchase from the World Tourism
Organisation at \url{https://www.eunwto.org/action/doSearch?ConceptID=2445&target=topic}.

Dengue notification data from Australia is publicly available from the Australian Department of Health at \url{http://www9.health.gov.au/cda/source/rpt_1_sel.cfm}.

Dengue notification data from Europe is publicly available from the European Centre for Disease Prevention and Control at \url{http://atlas.ecdc.europa.eu/public/index.aspx}.

Case-based dengue notifications for Queensland cannot be shared as it contains confidential information. The authors gained access to this data in accordance with Section 284 of the Public Health Act 2005.

\section*{Acknowledgments}
We would like to thank Frank de Hoog and Simon Dunstall for their constructive feedback which helped us to improve the model. We would also like to thank Queensland Health for providing dengue outbreak data. This work is part of the DiNeMo project.

% Either type in your references using
% \begin{thebibliography}{}
% \bibitem{}
% Text
% \end{thebibliography}
%
% or
%
% Compile your BiBTeX database using our plos2015.bst
% style file and paste the contents of your .bbl file
% here. See http://journals.plos.org/plosone/s/latex for 
% step-by-step instructions.
% 

%----------------------------------------------------------------------------------------
%	REFERENCE LIST
%----------------------------------------------------------------------------------------

\bibliographystyle{unsrt} 
\bibliography{DengueImportation}

\begin{thebibliography}{10}

\bibitem{Bloom2017}
David~E. Bloom, Steven Black, and Rino Rappuoli.
\newblock {Emerging infectious diseases: A proactive approach}.
\newblock {\em Proc. Natl. Acad. Sci.}, 114(16):4055--4059, 2017.

\bibitem{Brockmann2013}
Dirk Brockmann and Dirk Helbing.
\newblock {The hidden geometry of complex, network-driven contagion phenomena}.
\newblock {\em Science}, 342(6164):1337--1342, 2013.

\bibitem{Dorigatti2017}
Ilaria Dorigatti, Arran Hamlet, Ricardo Aguas, Lorenzo Cattarino, Anne Cori,
  Christl~A Donnelly, Tini Garske, Natsuko Imai, and Neil~M Ferguson.
\newblock {International risk of yellow fever spread from the ongoing outbreak
  in Brazil, December 2016 to May 2017}.
\newblock {\em Eurosurveillance}, 22(28):30572, 2017.

\bibitem{Guimera2005}
R.~Guimera, S.~Mossa, A.~Turtschi, and L.~A.~N. Amaral.
\newblock {The worldwide air transportation network: Anomalous centrality,
  community structure, and cities' global roles}.
\newblock {\em Proc. Natl. Acad. Sci.}, 102(22):7794--7799, 2005.

\bibitem{Huang2013}
Zhuojie Huang and Andrew~J Tatem.
\newblock {Global malaria connectivity through air travel}.
\newblock {\em Malar. J.}, 12(1):269, 2013.

\bibitem{Hufnagel2004}
L.~Hufnagel, D.~Brockmann, and T.~Geisel.
\newblock {Forecast and control of epidemics in a globalized world}.
\newblock {\em Proc. Natl. Acad. Sci.}, 101(42):15124--15129, 2004.

\bibitem{Tatem2012}
A.~J. Tatem, Z.~Huang, A.~Das, Q.~Qi, J.~Roth, and Y.~Qiu.
\newblock {Air travel and vector-borne disease movement}.
\newblock {\em Parasitology}, 139(14):1816--1830, 2012.

\bibitem{Bogoch2016}
Isaac~I Bogoch, Oliver~J Brady, Moritz U~G Kraemer, Matthew German, Marisa~I
  Creatore, Manisha~A Kulkarni, John~S Brownstein, Sumiko~R Mekaru, Simon~I
  Hay, Emily Groot, Alexander Watts, and Kamran Khan.
\newblock {Anticipating the international spread of Zika virus from Brazil}.
\newblock {\em Lancet}, 387(10016):335--336, 2016.

\bibitem{Bhatt2013}
Samir Bhatt, Peter~W. Gething, Oliver~J. Brady, Jane~P. Messina, Andrew~W.
  Farlow, Catherine~L. Moyes, John~M. Drake, John~S. Brownstein, Anne~G. Hoen,
  Osman Sankoh, Monica~F. Myers, Dylan~B. George, Thomas Jaenisch,
  G.~R.~William Wint, Cameron~P. Simmons, Thomas~W. Scott, Jeremy~J. Farrar,
  and Simon~I. Hay.
\newblock {The global distribution and burden of dengue}.
\newblock {\em Nature}, 496(7446):504--507, 2013.

\bibitem{Stanaway2016}
Jeffrey~D Stanaway, Donald~S Shepard, Eduardo~A Undurraga, Yara~A Halasa, Luc~E
  Coffeng, Oliver~J Brady, Simon~I Hay, Neeraj Bedi, Isabela~M Bensenor,
  Carlos~A Casta{\~{n}}eda-Orjuela, Ting-Wu Chuang, Katherine~B Gibney, Ziad~A
  Memish, Anwar Rafay, Kingsley~N Ukwaja, Naohiro Yonemoto, and Christopher J~L
  Murray.
\newblock {The global burden of dengue: An analysis from the Global Burden of
  Disease Study 2013}.
\newblock {\em Lancet Infect. Dis.}, 16(6):712--723, 2016.

\bibitem{Murray2013}
Natasha Evelyn~Anne Murray, Mikkel Quam, and Annelies Wilder-Smith.
\newblock {Epidemiology of dengue: Past, present and future prospects}.
\newblock {\em Clin. Epidemiol.}, 5:299--309, 2013.

\bibitem{WorldHealthOrganization2013}
{World Health Organization}.
\newblock {WHO heralds “new phase” in the fight against neglected tropical
  diseases}, 2013.
\newblock Available at:
  \url{http://www.who.int/mediacentre/news/releases/2013/ntds\_report\_20130116/en/}
  [Accessed 25/10/18].

\bibitem{Gardner2013}
Lauren Gardner and Sahotra Sarkar.
\newblock {A global airport-based risk model for the spread of dengue infection
  via the air transport network}.
\newblock {\em PLoS One}, 8(8):e72129, 2013.

\bibitem{Mackenzie2004}
John~S Mackenzie, Duane~J Gubler, and Lyle~R Petersen.
\newblock {Emerging flaviviruses: The spread and resurgence of Japanese
  encephalitis, West Nile and dengue viruses}.
\newblock {\em Nat. Med.}, 10(12):S98--S109, 2004.

\bibitem{Simmons2012}
Cameron~P. Simmons, Jeremy~J. Farrar, Nguyen {van Vinh Chau}, and Bridget
  Wills.
\newblock {Dengue}.
\newblock {\em N. Engl. J. Med.}, 366(15):1423--1432, 2012.

\bibitem{Chang2015}
Hsu P-S Chen C-D Lian I-B Chao D-Y Chang F-S, Tseng Y-T.
\newblock {Re-assess vector indices threshold as an early warning tool for
  predicting dengue epidemic in a dengue non-endemic country}.
\newblock {\em PLOS Negl Trop Dis.}, 9(9):e0004043, 2015.

\bibitem{Semenza2014}
Jan~C. Semenza, Bertrand Sudre, Jennifer Miniota, Massimiliano Rossi, Wei Hu,
  David Kossowsky, Jonathan~E. Suk, Wim {Van Bortel}, and Kamran Khan.
\newblock {International dispersal of dengue through air travel: Importation
  risk for Europe}.
\newblock {\em PLoS Negl. Trop. Dis.}, 8(12):e3278, 2014.

\bibitem{Gardner2015}
Lauren~M. Gardner and Sahotra Sarkar.
\newblock {Risk of dengue spread from the Philippines through international air
  travel}.
\newblock {\em Transp. Res. Rec. J. Transp. Res. Board}, 2501:25--30, 2015.

\bibitem{Wilder-Smith2014}
A.~Wilder-Smith, M.~Quam, O.~Sessions, J.~Rocklov, J.~Liu-Helmersson,
  L.~Franco, and K.~Khan.
\newblock {The 2012 dengue outbreak in Madeira: Exploring the origins}.
\newblock {\em Eurosurveillance}, 19(8), 2014.

\bibitem{Quam2015}
Mikkel~B. Quam, Kamran Khan, Jennifer Sears, Wei Hu, Joacim Rockl\"{o}v, and
  Annelies Wilder‐Smith.
\newblock {Estimating air travel–associated importations of dengue virus into
  Italy}.
\newblock {\em J. Travel Med.}, 22(3):186--193, 2015.

\bibitem{Massad2018}
Eduardo Massad, Marcos Amaku, Francisco Antonio~Bezerra Coutinho,
  Claudio~Jos{\'{e}} Struchiner, Marcelo~Nascimento Burattini, Kamran Khan,
  Jing Liu-Helmersson, Joacim Rockl{\"{o}}v, Moritz U.~G. Kraemer, and Annelies
  Wilder-Smith.
\newblock {Estimating the probability of dengue virus introduction and
  secondary autochthonous cases in Europe}.
\newblock {\em Sci. Rep.}, 8(1):4629, 2018.

\bibitem{Standish2010}
Katherine Standish, Guillermina Kuan, William Avil{\'{e}}s, Angel Balmaseda,
  and Eva Harris.
\newblock {High dengue case capture rate in four years of a cohort study in
  Nicaragua compared to national surveillance data}.
\newblock {\em PLoS Negl. Trop. Dis.}, 4(3):e633, 2010.

\bibitem{Kakkar2012}
M.~Kakkar.
\newblock {Dengue fever is massively under-reported in India, hampering our
  response}.
\newblock {\em BMJ}, 345(17):e8574, 2012.

\bibitem{Vong2012}
S.~Vong, S.~Goyet, S.~Ly, C.~Ngan, R.~Huy, V.~Duong, O.~Wichmann, G.~W. Letson,
  H.~S. Margolis, and P.~Buchy.
\newblock {Under-recognition and reporting of dengue in Cambodia: A
  capture–recapture analysis of the National Dengue Surveillance System}.
\newblock {\em Epidemiol. Infect.}, 140(03):491--499, 2012.

\bibitem{Wahyono2017}
T.~Y.~M. Wahyono, J.~Nealon, S.~Beucher, A.~Prayitno, A.~Moureau, S.~Nawawi,
  H.~Thabrany, and M.~Nadjib.
\newblock {Indonesian dengue burden estimates: Review of evidence by an expert
  panel}.
\newblock {\em Epidemiol. Infect.}, 145(11):2324--2329, 2017.

\bibitem{IAMAT}
{International Association for Medical Assistence to Travellers (IAMAT)}.
\newblock {International Association for Medical Assistence to Travellers},
  2018.
\newblock Available at: \url{https://www.iamat.org/} [Accessed 6/11/18].

\bibitem{GlobalHealthDataExchange2017}
{Global Burden of Disease Collaborative Network}.
\newblock {Global Burden of Disease Study 2016 (GBD 2016) Results}, 2017.
\newblock Available at: \url{http://ghdx.healthdata.org/gbd-results-tool}
  [Accessed 9/11/18].

\bibitem{Toan2015}
Nguyen~T. Toan, Stefania Rossi, Gabriella Prisco, Nicola Nante, and Simonetta
  Viviani.
\newblock {Dengue epidemiology in selected endemic countries: Factors
  influencing expansion factors as estimates of underreporting}.
\newblock {\em Trop. Med. Int. Heal.}, 20(7):840--863, 2015.

\bibitem{AirlineNetworkNewsandAnalysis2018}
{Airline Network News and Analysis}.
\newblock {Aviation database}, 2018.
\newblock Available at: \url{https://www.anna.aero/databases/} [Accessed
  1/12/18].

\bibitem{CivilAviationAdministrationofChina2016}
{Civil Aviation Administration of China}.
\newblock {2015 Civil Aviation Airport Throughput Ranking}, 2016.
\newblock Available at:
  \url{http://www.caac.gov.cn/XXGK/XXGK/TJSJ/201603/t20160331\_30105.html}
  [Accessed 1/12/18].

\bibitem{MinistryofLandInfrastructureTransportandTourism}
{Ministry of Land Infrastructure Transport and Tourism}.
\newblock {Ministry of Land, Infrastructure, Transport and Tourism,}, 2016.
\newblock Available at:
  \url{https://web.archive.org/web/20161021205147/http://www.mlit.go.jp/common/001141840.pdf}
  [Accessed 1/12/18].

\bibitem{MinistryofTransportationRepublicofIndonesia2018}
{Ministry of Transportation Republic of Indonesia}.
\newblock {Air transport traffic}, 2018.
\newblock Available at:
  \url{http://hubud.dephub.go.id/?en/llu/index/filter:bulan,0} [Accessed
  1/12/18].

\bibitem{AirportsCouncilInternational2015}
{Airports Council International}.
\newblock {North America airport traffic annual reports}, 2015.
\newblock Available at:
  \url{https://airportscouncil.org/intelligence/north-american-airport-traffic-reports/north-america-airport-traffic-annual-reports/}
  [Accessed 1/12/18].

\bibitem{DirectorateGeneralofCivilAviation-Kuwait2015}
{Directorate General of Civil Aviation - Kuwait}.
\newblock {Statistics}, 2015.
\newblock Available at:
  \url{https://www.dgca.gov.kw/en/civil-aviation/media-and-info/statistics}
  [Accessed 1/12/18].

\bibitem{EANA-NavegacionAereaArgentina2016}
{EANA - Navegaci{\'{o}}n A{\'{e}}rea Argentina}.
\newblock {Statistical reports}, 2016.
\newblock Available at: \url{https://www.eana.com.ar/estadisticas\#node-150}
  [Accessed 1/12/18].

\bibitem{AngkasaPuraII2015}
{Angkasa Pura II}.
\newblock {Innovating beyond excellence - 2015 annual report}, 2015.
\newblock Available at:
  \url{https://cms.angkasapura2.co.id/NUWEB\_PUBLIC\_FILES/angkasapura2/Annual\_25\_07\_2016\_\_09\_04\_23.pdf}
  [Accessed 1/12/18].

\bibitem{CivilAviationAuthorityofthePhilippines2015}
{Civil Aviation Authority of the Philippines}.
\newblock {Passenger movement}, 2015.
\newblock Available at:
  \url{https://data.gov.ph/dataset/civil-aviation-authority-philippines-passenger-movement}
  [Accessed 1/12/18].

\bibitem{InstitutodeEstadisticasdePuertoRico2015}
{Instituto de Estad{\'{i}}sticas de Puerto Rico}.
\newblock {Carga y passajeros a{\'{e}}reos y mar{\'{i}}timos}, 2015.
\newblock Available at:
  \url{http://www.estadisticas.gobierno.pr/iepr/Estadisticas/InventariodeEstad\%C3\%ADsticas/tabid/186/ctl/view_detail/mid/775/report_id/9485fbd8-efa5-4583-a8a4-a39c74eb846f/Default.aspx}
  [Accessed 1/12/18].

\bibitem{MalaysiaAirports2015}
{Malaysia Airports}.
\newblock {Connecting you seamlessly to the world - Annual report 2015}, 2015.
\newblock Available at: \url{http://mahb.listedcompany.com/misc/ar/ar2015.pdf}
  [Accessed 1/12/18].

\bibitem{Zambia:TransportDataPortal2016}
{Zambia: Transport Data Portal}.
\newblock {Air transport statistics Zambia}, 2016.
\newblock Available at:
  \url{http://zambiamtc.opendataforafrica.org/ZMATS2015/air-transport-statistics-zambia?airports=1000000-kenneth-kaunda-international-airport}
  [Accessed 1/12/18].

\bibitem{CaliforniaDepartmentofTransportation2016}
{California Department of Transportation}.
\newblock 2105:2014 air passenger and air cargo traffic activity report, 2016.
\newblock Available at:
  \url{http://www.dot.ca.gov/hq/planning/aeronaut/documents/statistics/15trafficgrowthrpt.pdf}
  [Accessed 1/12/18].

\bibitem{AirportsAuthoriyofIndia2016}
{Airports Authoriy of India}.
\newblock {Passengers}, 2016.
\newblock Available at:
  \url{https://www.aai.aero/sites/default/files/traffic-news/Mar2k15annex3.pdf}
  [Accessed 1/12/18].

\bibitem{AssociazioneItalianaGestoriAeroporti2016}
{Associazione Italiana Gestori Aeroporti}.
\newblock {Statistiche}, 2016.
\newblock Available at: \url{http://www.assaeroporti.com/statistiche/}
  [Accessed 1/12/18].

\bibitem{OfficeNationaldesAeroports2017}
{Office National des A{\'{e}}roports}.
\newblock {Passengers}, 2017.
\newblock Available at:
  \url{http://www.onda.ma/en/I-am-a-Professional/Companies/statistics/(offset)/30}
  [Accessed 1/12/18].

\bibitem{AirportsCompanySouthAfrica2018}
{Airports Company South Africa}.
\newblock {Statistcs}, 2018.
\newblock Available at: \url{http://www.airports.co.za/news/statistics}
  [Accessed 1/12/18].

\bibitem{EgyptianHoldingCompanyforAirportsandAirNavigation2015}
{Egyptian Holding Company for Airports and Air Navigation}.
\newblock {Statistics}, 2015.
\newblock Available at: \url{http://www.ehcaan.com/statistics.aspx} [Accessed
  1/12/18].

\bibitem{AirportsCouncilInternational2018}
{Airports Council International}.
\newblock {Passenger Summary}, 2018.
\newblock Available at:
  \url{https://aci.aero/data-centre/annual-traffic-data/passengers/2015-final-summary/}
  [Accessed 1/12/18].

\bibitem{DepartmentofHomeAffairs2017}
{Department of Home Affairs}.
\newblock {Overseas Arrivals and Departures}, 2017.
\newblock Available at:
  \url{https://data.gov.au/dataset/overseas-arrivals-and-departures} [Accessed
  22/11/18].

\bibitem{Chan2012}
Miranda Chan and Michael~A. Johansson.
\newblock {The incubation periods of dengue viruses}.
\newblock {\em PLoS One}, 7(11):e50972, 2012.

\bibitem{Chowell2007}
G.~Chowell, P.~Diaz-Due{\~{n}}as, J.C. Miller, A.~Alcazar-Velazco, J.M. Hyman,
  P.W. Fenimore, and C.~Castillo-Chavez.
\newblock {Estimation of the reproduction number of dengue fever from spatial
  epidemic data}.
\newblock {\em Math. Biosci.}, 208(2):571--589, 2007.

\bibitem{Gubler1995}
Duane~J. Gubler and Gary~G. Clark.
\newblock {Dengue/dengue hemorrhagic fever: The emergence of a global health
  problem}.
\newblock {\em Emerg. Infect. Dis.}, 1(2):55--57, 1995.

\bibitem{Messenger2007}
McCann~D Messenger~JC, Lee~S.
\newblock {\em {Working time around the world: Trends in working hours, laws,
  and policies in a global comparative perspective.}}
\newblock 1st ed. London: Routledge, 2007.

\bibitem{CentralIntelligenceAgency2017}
{Central Intelligence Agency}.
\newblock {The World Factbook}, 2017.
\newblock Available at:
  \url{https://www.cia.gov/library/publications/the-world-factbook/fields/2177.html{\#}aq}
  [Accessed 12/11/18].

\bibitem{WorldTourismOrganisation2018}
{World Tourism Organisation}.
\newblock {All Countries: Inbound Tourism: Arrivals by mode of transport 1995 -
  2017}, 2018.
\newblock Available at:
  \url{https://www.e-unwto.org/doi/abs/10.5555/unwtotfb0000271519952017201809}
  [Accessed 12/11/18].

\bibitem{Sobol2001}
Sobol IM.
\newblock {Global sensitivity indices for nonlinear mathematical models and
  their Monte Carlo estimates}.
\newblock {\em Math Comput Simulat.}, 55:271--280, 2001.

\bibitem{Herman2017}
Usher~W Herman~J.
\newblock {SALib: An open-source Python library for sensitivity analysis}.
\newblock {\em The Journal of Open Source Software}, 2(9):97, 2017.

\bibitem{Morrison2008}
Amy~C Morrison, Emily Zielinski-Gutierrez, Thomas~W Scott, and Ronald
  Rosenberg.
\newblock {Defining challenges and proposing solutions for control of the virus
  vector Aedes aegypti}.
\newblock {\em PLoS Med.}, 5(3):e68, 2008.

\bibitem{Beebe2009}
Nigel~W. Beebe, Robert~D. Cooper, Pipi Mottram, and Anthony~W. Sweeney.
\newblock {Australia's dengue risk driven by human adaptation to climate
  change}.
\newblock {\em PLoS Negl. Trop. Dis.}, 3(5):e429, 2009.

\bibitem{Hahn2016}
Micah~B. Hahn, Rebecca~J. Eisen, Lars Eisen, Karen~A. Boegler, Chester~G.
  Moore, Janet McAllister, Harry~M. Savage, and John-Paul Mutebi.
\newblock {Reported Distribution of Aedes (Stegomyia) aegypti and Aedes
  (Stegomyia) albopictus in the United States, 1995-2016 (Diptera: Culicidae)}.
\newblock {\em J. Med. Entomol.}, 53(5):1169--1175, 2016.

\bibitem{VanDodewaard2015}
Caitlin~A.M. {Van Dodewaard} and Stephanie~L. Richards.
\newblock {Trends in dengue cases imported into the United States from Pan
  America 2001--2012}.
\newblock {\em Environ. Health Insights}, 9:33--40, 2015.

\bibitem{Amarakoon2008}
Amarakoon, Dharmaratne, Anthony Chen, Sam Rawlins, Dave~D. Chadee, Michael
  Taylor, and Roxann Stennett.
\newblock {Dengue epidemics in the Caribbean--temperature indices to gauge the
  potential for onset of dengue}.
\newblock {\em Mitig. Adapt. Strateg. Glob. Chang.}, 13(4):341--357, 2008.

\bibitem{Vasquez2018}
Victor Vasquez, Elie Haddad, Alice Perignon, St{\'{e}}phane Jaureguiberry,
  S{\'{e}}gol{\`{e}}ne Brichler, Isabelle Leparc-Goffart, and Eric Caumes.
\newblock {Dengue, chikungunya, and Zika virus infections imported to Paris
  between 2009 and 2016: Characteristics and correlation with outbreaks in the
  French overseas territories of Guadeloupe and Martinique}.
\newblock {\em Int. J. Infect. Dis.}, 72:34--39, 2018.

\bibitem{LaRuche2013}
Guy {La Ruche}, Dominique Dejour-Salamanca, Pascale Bernillon, Isabelle
  Leparc-Goffart, Martine Ledrans, Alexis Armengaud, Monique Debruyne,
  G{\'{e}}rard-Antoine Denoyel, and S{\'{e}}gol{\`{e}}ne Brichler.
\newblock {Capture–recapture method for estimating annual incidence of
  imported dengue, France, 2007--2010}.
\newblock {\em Emerg. Infect. Dis.}, 19(11):1740--1748, 2013.

\bibitem{Warrilow2012}
David Warrilow, Judith~A. Northill, and Alyssa~T. Pyke.
\newblock {Sources of dengue viruses imported into Queensland, Australia,
  2002--2010}.
\newblock {\em Emerg. Infect. Dis.}, 18(11):1850--1857, 2012.

\bibitem{TheFijianGovernment2014}
{The Fijian Government}.
\newblock {Minimal risk of dengue fever in major tourism areas of Fiji}, 2014.
\newblock Available at:
  \url{http://www.fiji.gov.fj/Media-Center/Press-Releases/MINIMAL-RISK-OF-DENGUE-FEVER-IN-MAJOR-TOURISM-AREA.aspx}
  [Accessed 8/6/18].

\bibitem{QueenslandHealth2015}
{Queensland Health}.
\newblock {Queensland dengue management plan 2015--2020}, 2015.
\newblock Available at:
  \url{https://www.health.qld.gov.au/\_\_data/assets/pdf\_file/0022/444433/dengue-mgt-plan.pdf}
  [Accessed 29/10/18].

\bibitem{Hayden2015}
Mary~H. Hayden, Jamie~L. Cavanaugh, Christopher Tittel, Melinda Butterworth,
  Steven Haenchen, Katherine Dickinson, Andrew~J. Monaghan, and Kacey~C. Ernst.
\newblock {Post outbreak review: Dengue preparedness and response in Key West,
  Florida}.
\newblock {\em Am. J. Trop. Med. Hyg.}, 93(2):397--400, 2015.

\bibitem{healthmap}
Centers for Disease~Control and Prevention.
\newblock {Dengue map}, 2019.
\newblock Available at: \url{https://www.healthmap.org/dengue/en/} [Accessed
  31/07/19].

\end{thebibliography}

%----------------------------------------------------------------------------------------

	\onecolumn
	\setcounter{table}{0}
	\renewcommand{\thetable}{S\arabic{table}}
	
	\setcounter{figure}{0}
	\renewcommand{\thefigure}{S\arabic{figure}}
	\section*{Supporting information}

\begin{table*}[h]
	\caption{\normalfont\textbf{List of airport abbreviations}}
	\centering
	\scriptsize
	\begin{tabular}{p{1cm}p{3cm}p{3cm}|p{1cm}p{3cm}p{3cm}}
		IATA 3-Letter Code & Name & City (Country/State)& IATA 3-Letter Code & Name & City (Country/State)\\
		\midrule
		AEP& Jorge Newbery Airport &Buenos Aires (Argentina)&LAX&Los Angeles International Airport& Los Angeles (California)\\
		BKK&Suvarnabhumi Airport&Bangkok (Thailand)&LHR&Heathrow Airport&London (UK)\\
		BNE&Brisbane Airport &Brisbane (Queensland)&MAD&Adolfo Su\'{a}rez Madrid–Barajas Airport &Madrid (Spain)\\
		BOM & Chhatrapati Shivaji International Airport &Mumbai (India) &MCO&Orlando International Airport &Orlando (Florida)\\
		CDG&Charles de Gaulle Airport &Paris (France)&MEX&Mexico City International Airport &Mexico City (Mexico)\\
		COK&Cochin International Airport &Kochi (India)&MIA&Miami International Airport &Miami (Florida)\\
		CUN &Canc\'{u}n International Airport &Canc\'{u}n (Mexico)&MNL&Ninoy Aquino International Airport &Manila (Philippines)\\
		DEL&Indira Gandhi International Airport &New Delhi (India)&MTY&Monterrey International Airport &Apodaca (Mexico)\\
		DFW&Dallas/Fort Worth International Airport &Dallas (Texas)&MXP&Milan Malpensa Airport &Milan (Italy)\\
		DPS&Ngurah Rai International Airport &Denpasar (Indonesia)&NRT&Narita International Airport&Tokyo (Japan)\\
		DXB&Dubai International Airport &Dubai (UAE)&ORY&Paris Orly Airport &Paris (France)\\
		EZE&Ministro Pistarini International Airport &Buenos Aires (Argentina)&PER&Perth Airport &Perth (Western Australia)\\
		FDF&Martinique Aim\'{e} C\'{e}saire International Airport& Forte-de-France (Martinique)&PTP&Pointe-\`{a}-Pitre International Airport& Pointe-\`{a}-Pitre (Guadeloupe)\\
		FLL&Fort Lauderdale–Hollywood International Airport &Miami (Florida)&PUJ&Punta Cana International Airport& Punta Cana (Dominican Republic)\\
		GDL&Miguel Hidalgo y Costilla Guadalajara International Airport& Guadalajara (Mexico)&SAL&Monse\~{n}or \'{O}scar Arnulfo Romero International Airport & San Salvador (El Salvador)\\
		GRU&S\~{a}o Paulo International Airport& S\~{a}o Paulo (Brazil)&SDQ&Las Am\'{e}ricas International Airport& Punta Caucedo (Dominican Republic)\\
		ICN&Incheon International Airport &Seoul (South Korea)&SFO&San Francisco International Airport& San Francisco (California)\\
		IAH&George Bush Intercontinental Airport &Houston (Texas)&SJU&Luis Mu\~{n}oz Mar\'{i}n International Airport& San Juan (Puerto Rico)\\		
		JFK&John F. Kennedy International Airport& New York City (New York)&STI&Cibao International Airport&Santiago de los Caballeros (Dominican Republic)\\
		KBL&Hamid Karzai International Airport&Kabul (Afghanistan)&TPE&Taiwan Taoyuan International Airport& Taipei (Taiwan)\\
		\bottomrule
	\end{tabular}
	\label{tab:abbreviations}
\end{table*}

\clearpage
\begin{table*}[h]
	\caption{\normalfont\textbf{List of non-endemic countries where IATA data is inaccurate}}
	\centering
	\scriptsize
	\begin{tabular}{llll}
		\midrule
		Algeria&Bahrain&Bonaire, Saint Eustatius \& Saba&Bulgaria\\
		Central African Republic&Croatia&Egypt&Federated States of Micronesia\\
		Finland&Germany&Guinea-Bissau&Greece\\
		Hungary&Iceland&Iran&Israel\\
		Malawi&Morocco&Netherlands&Russian Federation\\
		Serbia&Slovenia&South Africa&South Korea\\
		Tanzania&Togo&The Gambia&Tunisia\\
		Turkey&Uganda&Ukraine&Zambia\\
		\bottomrule
	\end{tabular}
	\label{tab:excluded_countries}
\end{table*}

\vspace*{1cm}
{\scriptsize
	\begin{longtable}{p{2cm}lp{1.3cm}p{2cm}lp{1.3cm}p{2cm}lp{1.3cm}}
		\caption{\normalfont\textbf{List of countries indicating whether dengue vectors are present and whether the country is endemic. Information about endemicity was obtained from~\cite{healthmap}. Information about vector presence was obtained from~\cite{IAMAT}.}}\label{tab:endemic_vector}\\
		Country & Endemic & Vector presence & Country & Endemic & Vector presence & Country & Endemic & Vector presence\\
		\midrule
		\endfirsthead
		\endhead
		Country & Endemic & Vector presence & Country & Endemic & Vector presence & Country & Endemic & Vector presence\\
		\midrule
		\endhead 
		Afghanistan & no & yes & Ghana & no & yes & Pakistan & yes & yes \\  
		Albania & no & yes & Gibraltar & no & yes & Palau & yes & yes \\  
		Algeria & no & yes & Greece & no & yes & Palestine & no & yes \\  
		American Samoa & no & yes & Greenland & no & no & Panama & yes & yes \\  
		Andorra & no & no & Grenada & yes & yes & Papua New Guinea & yes & yes \\  
		Angola & yes & yes & Guadeloupe & yes & yes & Paraguay & yes & yes \\  
		Anguilla & yes & yes & Guam & no & no & Peru & yes & yes \\  
		Antarctica & no & no & Guatemala & yes & yes & Philippines & yes & yes \\  
		Antigua and Barbuda & yes & yes & Guinea & yes & yes & Poland & no & no \\  
		Argentina & yes & yes & Guinea-Bissau & no & yes & Portugal & no & yes \\  
		Armenia & no & yes & Guyana & yes & yes & Puerto Rico & yes & yes \\  
		Aruba & yes & yes & Haiti & yes & yes & Qatar & no & no \\  
		Australia & no & yes & Honduras & yes & yes & Reunion & yes & yes \\  
		Austria & no & yes & Hong Kong & yes & yes & Romania & no & yes \\  
		Azerbaijan & no & no & Hungary & no & yes & Russian Federation & no & yes \\  
		Bahrain & no & no & Iceland & no & no & Rwanda & no & yes \\  
		Bangladesh & yes & yes & India & yes & yes & Saint Helena & no & no \\  
		Barbados & yes & yes & Indonesia & yes & yes & Saint Kitts and Nevis & yes & yes \\  
		Belarus & no & no & Inner Hebrides & no & no & Saint Lucia & yes & yes \\  
		Belgium & no & yes & Iran & no & no & Saint Pierre and Miquelon & no & no \\  
		Belize & yes & yes & Iraq & no & no & Saint Vincent and the Grenadines & yes & yes \\  
		Benin & no & yes & Ireland & no & no & Samoa & no & yes \\  
		Bermuda & no & yes & Israel & no & yes & Sao Tome and Principe & no & no \\  
		Bhutan & yes & yes & Italy & no & yes & Saudi Arabia & yes & yes \\  
		Bolivia & yes & yes & Jamaica & yes & yes & Senegal & yes & yes \\  
		Bonaire, Saint Eustatius \& Saba & no & yes & Japan & no & no & Serbia & no & yes \\  
		Bosnia and Herzegovina & no & yes & Jordan & no & yes & Seychelles & yes & yes \\  
		Botswana & no & no & Kazakhstan & no & no & Sierra Leone & yes & yes \\  
		Brazil & yes & yes & Kenya & yes & yes & Singapore & yes & yes \\  
		Brunei & yes & yes & Kiribati & no & yes & Sint Maarten & no & yes \\  
		Bulgaria & no & yes & Kuwait & no & no & Slovakia & no & yes \\  
		Burkina Faso & yes & yes & Kyrgyzstan & no & no & Slovenia & no & yes \\  
		Burundi & no & yes & Laos & yes & yes & Solomon Islands & yes & yes \\  
		Cambodia & yes & yes & Latvia & no & no & Somalia & yes & yes \\  
		Cameroon & yes & yes & Lebanon & no & yes & South Africa & no & no \\  
		Canada & no & no & Lesotho & no & no & South Korea & no & no \\  
		Cape Verde & yes & yes & Liberia & no & yes & South Sudan & yes & yes \\  
		Cayman Islands & yes & yes & Libya & no & no & Spain & no & yes \\  
		Central African Republic & no & yes & Liechtenstein & no & no & Sri Lanka & yes & yes \\  
		Chad & no & yes & Lithuania & no & no & Sudan & yes & yes \\  
		Channel Islands & no & no & Luxembourg & no & no & Suriname & yes & yes \\  
		Chile & no & no & Macau & yes & yes & Swaziland & no & no \\  
		China & yes & yes & Macedonia & no & yes & Sweden & no & no \\  
		Christmas Island & no & no & Madagascar & yes & yes & Switzerland & no & yes \\  
		Cocos (Keeling) Islands & no & no & Malawi & no & yes & Syria & no & yes \\  
		Colombia & yes & yes & Malaysia & yes & yes & Taiwan & yes & yes \\  
		Comoros & yes & yes & Maldives & no & yes & Tajikistan & no & no \\  
		Congo & no & yes & Mali & yes & yes & Tanzania & no & yes \\  
		Cook Islands & no & yes & Malta & no & yes & Thailand & yes & yes \\  
		Costa Rica & yes & yes & Marshall Islands & no & yes & The Bahamas & yes & yes \\  
		Cote d'Ivoire & yes & yes & Martinique & yes & yes & The Gambia & no & yes \\  
		Croatia & no & yes & Mauritania & no & yes & Timor-Leste & yes & yes \\  
		Cuba & yes & yes & Mauritius & yes & yes & Togo & no & yes \\  
		Curacao & no & yes & Mayotte & yes & yes & Tonga & no & yes \\  
		Cyprus & no & no & Mexico & yes & yes & Trinidad and Tobago & yes & yes \\  
		Czech Republic & no & yes & Moldova & no & no & Tunisia & no & no \\  
		Democratic Republic of the Congo & yes & yes & Monaco & no & yes & Turkey & no & yes \\  
		Denmark & no & no & Mongolia & no & no & Turkmenistan & no & no \\  
		Djibouti & yes & yes & Montenegro & no & yes & Turks and Caicos Islands & yes & yes \\  
		Dominica & yes & yes & Montserrat & yes & yes & Tuvalu & no & yes \\  
		Dominican Republic & yes & yes & Morocco & no & no & Uganda & no & yes \\  
		Ecuador & yes & yes & Mozambique & yes & yes & Ukraine & no & no \\  
		Egypt & no & yes & Myanmar & yes & yes & United Arab Emirates & yes & yes \\  
		El Salvador & yes & yes & Namibia & no & yes & United Kingdom & no & no \\  
		Equatorial Guinea & yes & yes & Nauru & no & yes & United States & no & yes \\  
		Eritrea & yes & yes & Nepal & yes & yes & United States Minor Outlying Islands & no & no \\  
		Estonia & no & no & Netherlands & no & yes & Uruguay & no & no \\  
		Ethiopia & yes & yes & New Caledonia & yes & yes & Uzbekistan & no & no \\  
		Falkland Islands & no & no & New Zealand & no & no & Vanuatu & yes & yes \\  
		Federated States of Micronesia & no & yes & Nicaragua & yes & yes & Venezuela & yes & yes \\  
		Fiji & no & yes & Niger & no & yes & Vietnam & yes & yes \\  
		Finland & no & no & Nigeria & yes & yes & Virgin Islands & yes & yes \\  
		France & no & yes & Niue & no & yes & Wallis and Futuna Islands & no & yes \\  
		French Guiana & yes & yes & Norfolk Island & no & no & Western Sahara & no & no \\  
		French Polynesia & no & yes & North Korea & no & no & Yemen & yes & yes \\  
		Gabon & yes & yes & Northern Mariana Islands & no & yes & Zambia & no & yes \\  
		Georgia & no & yes & Norway & no & no & Zimbabwe & no & yes \\  
		Germany & no & yes & Oman & yes & yes & \  & \  & \  \\  
		\bottomrule
\end{longtable}}

\clearpage
%\begin{table*}[b]
%	\centering
{\scriptsize
	\begin{longtable}{p{1cm}p{1.5cm}p{1.5cm}p{5cm}p{3cm}p{2cm}}
		\caption{\normalfont\textbf{Annual estimated imported dengue cases per airport}}\label{tab:annual_imported}\\
		Code & Imported cases 2011 & Imported cases 2015& Name & City& Country/State  \\ 
		\midrule
		\endfirsthead
		\endhead
		Code & Imported cases 2011 & Imported cases 2015 & Name & City & Country/State\\ 
		\midrule
		\endhead 
		MIA & 2413 & 2547 & Miami International & Miami & US/Florida \\ 
		LAX & 1518 & 1871 & Los Angeles Intl & Los Angeles & US/California \\ 
		CDG & 941 & 1227 & Charles De Gaulle & Paris-De Gaulle & France/\^{I}le-de-France \\ 
		SFO & 831 & 1166 & San Francisco Intl & San Francisco & US/California \\ 
		MCO & 822 & 1036 & Orlando Intl & Orlando & US/Florida \\ 
		FLL & 792 & 970 & Ft Lauderdale Intl & Fort Lauderdale & US/Florida \\ 
		ORY & 652 & 788 & Orly & Paris-Orly & France/\^{I}le-de-France \\ 
		IAH & 599 & 810 & George Bush Intercontinental & Houston-Intercontinental & US/Texas \\ 
		EZE & 586 & 648 & Ministro Pistarini & Buenos Aires & Argentina/Autonomous City of Buenos Aires \\ 
		MAD & 385 & 439 & Adolfo Suarez-Barajas & Madrid & Spain/Community of Madrid \\ 
		DFW & 381 & 478 & Dallas/Ft Worth Intl & Dallas/Fort Worth & US/Texas \\ 
		BNE & 379 & 529 & Brisbane Intl & Brisbane & Australia/Queensland \\ 
		MXP & 356 & 408 & Malpensa & Milan-Malpensa & Italy/Lombardy \\ 
		FCO & 343 & 424 & Fiumicino & Rome-Da Vinci & Italy/Lazio \\ 
		AEP & 258 & 225 & Jorge Newbery & Buenos Aires-Newbery & Argentina/Autonomous City of Buenos Aires \\ 
		MLE & 204 & 214 & Ibrahim Nasir International & Male & Maldives/Mal\'{e} \\ 
		POS & 198 & 150 & Piarco International & Port of Spain & Trinidad and Tobago/Port of Spain \\ 
		TPA & 160 & 224 & Tampa International & Tampa & US/Florida \\ 
		SXM & 159 & 184 & Prinses Juliana International & St. Maarten & Sint Maarten \\ 
		BCN & 148 & 207 & Barcelona & Barcelona & Spain/Catalonia \\ 
		CUR & 147 & 173 & Hato International & Curacao & Curacao \\ 
		SAN & 142 & 181 & San Diego International Airport & San Diego & US/California \\ 
		HNL & 123 & 153 & Honolulu Intl & Honolulu/Oahu & US/Hawaii \\ 
		BEY & 122 & 162 & Rafic Hariri International & Beirut & Lebanon/Beirut \\ 
		KBL & 109 & 155 & Kabul International & Kabul & Afghanistan/Kabul \\ 
		ACC & 97 & 126 & Kotoka International & Accra & Ghana/Greater Accra \\ 
		SJC & 88 & 99 & San Jose Municipal & San Jose & US/California \\ 
		SAT & 80 & 127 & San Antonio Intl & San Antonio & US/Texas \\ 
		VCE & 75 & 108 & Marco Polo & Venice & Italy/Veneto \\ 
		AUS & 71 & 123 & Austin-Bergstrom International Airport & Austin & US/Texas \\ 
		SMF & 70 & 94 & Sacramento International & Sacramento & US/California \\ 
		OAK & 65 & 65 & Metro Oakland Intl & Oakland & US/California \\ 
		LYS & 56 & 72 & Satolas & Lyon & France/Auvergne-Rh\^{o}ne-Alpes \\ 
		OOL & 55 & 126 & Coolangatta & Gold Coast & Australia/Queensland \\ 
		JAX & 53 & 74 & Jacksonville Intl & Jacksonville & US/Florida \\ 
		MRS & 53 & 69 & Marignane & Marseille & France/Provence-Alpes-C\^{o}te d'Azur \\ 
		NCE & 53 & 71 & Cote D'Azur & Nice & France/Provence-Alpes-C\^{o}te d'Azur \\ 
		BLQ & 49 & 59 & Guglielmo Marconi & Bologna & Italy/Emilia-Romagna \\ 
		COR & 46 & 78 & Pajas Blancas & C\'{o}rdoba & Argentina/C\'{o}rdoba \\ 
		TLS & 42 & 53 & Blagnac & Toulouse & France/Occitanie \\ 
		COO & 39 & 66 & Cadjehoun & Cotonou & Benin/Littoral \\ 
		ONT & 39 & 43 & Ontario Intl & Ontario & US/California \\ 
		SAH & 38 & 11 & Sana'a International & Sana'a & Yemen/Sana'a \\ 
		LIN & 38 & 56 & Linate & Milan-Linate & Italy/Lombardy \\ 
		FAT & 37 & 42 & Fresno Yosemite International & Fresno & US/California \\ 
		SNA & 37 & 58 & John Wayne Airport & Orange County & US/California \\ 
		DLA & 33 & 60 & Douala International & Douala & Cameroon/Littoral \\ 
		KGL & 29 & 48 & Kigali International Airport & Kigali & Rwanda/Kigali \\ 
		PNS & 27 & 31 & Pensacola International & Pensacola & US/Florida \\ 
		BOD & 24 & 33 & Merignac & Bordeaux & France/Nouvelle-Aquitaine \\ 
		CAY & 24 & 38 & Felix Eboue & Cayenne & France/French Guiana \\ 
		NAN & 23 & 27 & Nadi International & Nadi & Fiji/Ba \\ 
		PBM & 22 & 34 & Johan A. Pengel Intl & Paramaribo & Suriname/Paramaribo \\ 
		CNS & 22 & 37 & Cairns International & Cairns & Australia/Queensland \\ 
		ELP & 22 & 25 & El Paso Intl & El Paso & US/Texas \\ 
		NTE & 21 & 23 & Chateau Bougon & Nantes & France/Pays de la Loire \\ 
		FLR & 21 & 26 & Peretola & Florence & Italy/Tuscany \\ 
		BUR & 18 & 9 & Hollywood-Burbank & Burbank & US/California \\ 
		HOU & 18 & 38 & William P Hobby & Houston-Hobby & US/Texas \\ 
		PPG & 17 & 18 & Pago Pago Intl & Pago Pago & American Samoa/Maoputasi County \\ 
		TRN & 17 & 20 & Citta Di Torino & Turin & Italy/Piedmont \\ 
		ROB & 17 & 17 & Roberts Intl & Monrovia-Roberts & Liberia \\ 
		MFE & 15 & 14 & Miller International & McAllen & US/Texas \\ 
		PBI & 15 & 23 & Palm Beach Intl & West Palm Beach & US/Florida \\ 
		LGB & 15 & 13 & Long Beach Municipal & Long Beach & US/California \\ 
		BZV & 15 & 29 & Maya Maya & Brazzaville & Congo/Brazzaville \\ 
		NAP & 14 & 16 & Capodichino & Naples & Italy/Campania \\ 
		TLH & 14 & 21 & Tallahassee International & Tallahassee & US/Florida \\ 
		MDZ & 13 & 20 & El Plumerillo & Mendoza & Argentina/Mendoza \\ 
		ROS & 13 & 25 & Islas Malvinas & Rosario & Argentina/Santa Fe \\ 
		CRP & 13 & 15 & Corpus Christi Intl & Corpus Christi & US/Texas \\ 
		RSW & 13 & 14 & Southwest Florida International & Fort Myers & US/Florida \\ 
		PMI & 13 & 15 & Palma De Mallorca & Palma de Mallorca & Spain/Balearic Islands \\ 
		MPL & 13 & 18 & Mediterranee & Montpellier & France/Occitanie \\ 
		VPS & 12 & 14 & Destin-Ft Walton Beach Airport & Destin-Ft Walton Beach & US/Florida \\ 
		VRN & 12 & 8 & Verona & Verona & Italy/Veneto\\ 
		FGI & 11 & 20 & Fagali'I & Apia & Samoa/Tuamasaga \\ 
		GNV & 11 & 16 & J R Alison Regional Municipal & Gainesville & US/Florida \\ 
		OGG & 11 & 13 & Kahului & Kahului/Maui & US/Hawaii \\ 
		VLC & 11 & 16 & Valencia Airport & Valencia & Spain/Valencian Community \\ 
		CTA & 11 & 11 & Fontanarossa & Catania & Italy/Sicily \\ 
		BJM & 10 & 17 & Bujumbura Intl & Bujumbura & Burundi/Bujumbura Mairie \\ 
		TAB & 10 & 8 & ANR Robinson International & Tobago & Trinidad and Tobago \\ 
		AGP & 10 & 15 & Malaga Airport & Malaga & Spain/Andalusia \\ 
		ADE & 10 & 3 & Aden International & Aden & Yemen/Aden \\ 
		HRE & 10 & 22 & Harare International & Harare & Zimbabwe/Harare \\ 
		PNR & 9 & 19 & Pointe Noire & Pointe Noire & Congo \\ 
		PUF & 9 & 10 & Uzein & Pau & France/Nouvelle-Aquitaine \\ 
		BIO & 9 & 15 & Bilbao Airport & Bilbao & Spain/Basque Autonomous Community \\ 
		GOA & 9 & 10 & Cristoforo Colombo & Genoa & Italy/Liguria \\ 
		LPA & 9 & 14 & Gran Canaria & Gran Canaria & Spain/Canary Islands \\ 
		GRK & 9 & 11 & Regional/R.Gray AAF & Killeen/Fort Hood & US/Texas \\ 
		WDH & 9 & 10 & Windhoek Intl & Windhoek & Namibia/Khomas \\ 
		MAF & 9 & 13 & Midland-Odessa Regl & Midland/Odessa & US/Texas \\ 
		TRW & 9 & 11 & Bonriki International & Tarawa & Kiribati \\ 
		TSV & 9 & 12 & Townsville International & Townsville & Australia/Queensland \\ 
		PSP & 9 & 11 & Palm Springs Muni & Palm Springs & US/California \\ 
		MLH & 9 & 10 & Euroairport & Mulhouse/Basel & France/Grand Est \\ 
		BES & 8 & 11 & Bretagne & Brest & France/Brittany \\ 
		NKC & 8 & 13 & Nouakchott & Nouakchott & Mauritania/Nouakchott \\ 
		BRC & 8 & 12 & San Carlos Bariloche International & San Carlos Bariloche & Argentina/R\'{i}o Negro \\ 
		LBB & 8 & 10 & Preston Smith Intl & Lubbock & US/Texas \\ 
		AMA & 8 & 10 & Rick Husband Intl & Amarillo & US/Texas \\ 
		EYW & 7 & 12 & Key West Intl & Key West & US/Florida \\ 
		NDJ & 7 & 16 & Ndjamena & N'Djamena & Chad/N'Djamena \\ 
		DAL & 7 & 22 & Dallas Love Field & Dallas-Love & US/Texas \\ 
		ECP & 5 & 11 & Northwest Florida Beaches International Airport & Panama City & US/Florida \\ 
		NIM & 5 & 12 & Diori Hamani International Airport & Niamey & Niger/Niamey \\ 
		SPN & 3 & 11 & Saipan International & Saipan & Northern Mariana Islands \\ 
		SFB & 0 & 10 & Sanford International & Orlando-Sanford & US/Florida \\ 
		\bottomrule
\end{longtable}}
%\end{table*}

\vspace*{1cm}
\begin{table}[!ht]
	\centering
	\scriptsize
	\caption{\bf The ten routes with the highest predicted number of dengue-infected passengers with final destinations in non-endemic countries with vector presence. }
	\begin{tabular}{llll}
		\toprule
		Orig. & Dest. & Pax & Month\\
		\midrule
		SJU (Puerto Rico)&MCO (Florida)&52&Jul\\ 
		FDF (Martinique)&ORY (France)&34&Aug\\ 
		CUN (Mexico)&MIA (Florida)&32&Aug\\
		SDQ (Dominican Republic)&MIA (Florida)&30&Aug\\
		CCS (Venezuela)&MIA (Florida)&28&Aug\\
		GDL (Mexico)&LAX (California)&27&Aug\\
		SJU (Puerto Rico)&FLL (Florida)&25&Jul\\ 
		PUJ (Dominican Republic)&MIA (Florida)&24&Jul\\
		MNL (Philippines)&LAX (California)&23&Jul\\
		SJU (Puerto Rico)&MIA (Florida)&23&Jul\\
		\bottomrule
	\end{tabular}
	\label{tab:routes2011}
\end{table}

\begin{table}[h]
	\centering
	\scriptsize
	\caption{\normalfont\textbf{The ten routes with the highest predicted number of dengue-infected passengers who continue to travel to non-endemic regions.} The table lists the direct routes with the highest predicted volume of dengue-infected passengers who continue to travel to non-endemic regions irrespective of vector presence. The last column records the month during which the highest number of infected passengers are predicted.}
	\begin{tabular}{p{2cm}p{2cm}p{1cm}p{1cm}|p{2cm}p{2cm}p{1cm}p{1cm}}
		\multicolumn{4}{c}{2011}&\multicolumn{4}{c}{2015}\\
		Origin & Destination & Pax & Month&Origin & Destination & Pax & Month\\
		\midrule
		BOM&DXB&108&Jul& BOM & DXB & 142 & Aug\\
		CUN&MEX&86&Aug&DEL & DXB & 97 & Aug \\
		DPS&PER&77&Jan&CUN & MEX & 75 & Aug\\
		SDQ&JFK&76&Aug&COK&DXB&72&Aug\\
		STI&JFK&76&Aug&DPS & PER & 65 & Jan\\
		DEL&DXB&72&Jul&MAA & DXB & 59 & Aug\\
		MNL&ICN&71&Aug&MNL & ICN & 95 & Aug\\
		DEL&LHR&65&Aug&HYD & DXB & 55 & Aug\\
		MNL&NRT&65&Jul&SJU & JFK & 57 & Aug\\
		MTY&MEX&62&Sep&DEL & LHR & 59 & Aug\\
		\bottomrule
	\end{tabular}
	\label{tab:routesS}
\end{table}

\begin{table}[!ht]
	\centering
	\scriptsize
	\centering
	\caption{\bf Yearly and seasonal reporting rates of imported cases in 2011.}
	\begin{tabular}{lccccc}
		\toprule
		&Dec-Feb & Mar-May & Jun-Aug & Sep-Nov & Yearly\\ 
		\midrule
		Queensland &38.4 &25.2 &13.7&18.9&24.3  \\
		Italy&5.7&4&1.9&7.3&4.4  \\
		France &2.2&3.1&4.8&1.3&3  \\
		Florida &1.2&0.5&1&2.5&1.3 \\
		\bottomrule
	\end{tabular}
	\label{tab:reportingRates2011}
\end{table}

\clearpage
\vspace*{5mm}
\begin{figure}[h]
	\centering
	\includegraphics[width=0.5\textwidth]{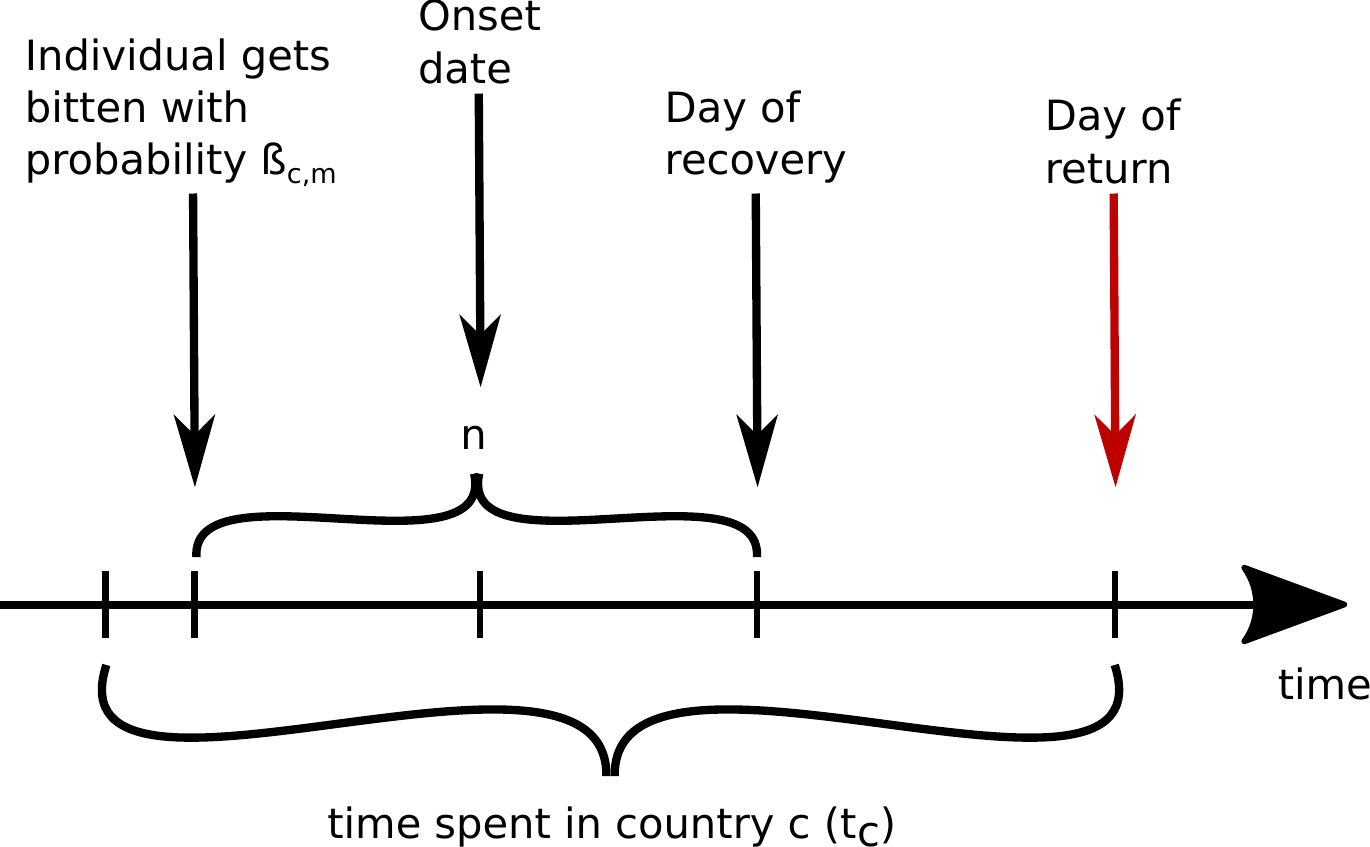}
	\caption{{\bf Illustration of the possibility of recovery before return.} \normalfont If an individual gets infected with dengue while overseas, but recovers before returning to region $r$, the individual cannot infect other people in region $r$.}
	\label{fig:diagram}
\end{figure}

\vspace*{5mm}
\begin{figure}[h]
	\centering
	\includegraphics[width=0.5\textwidth]{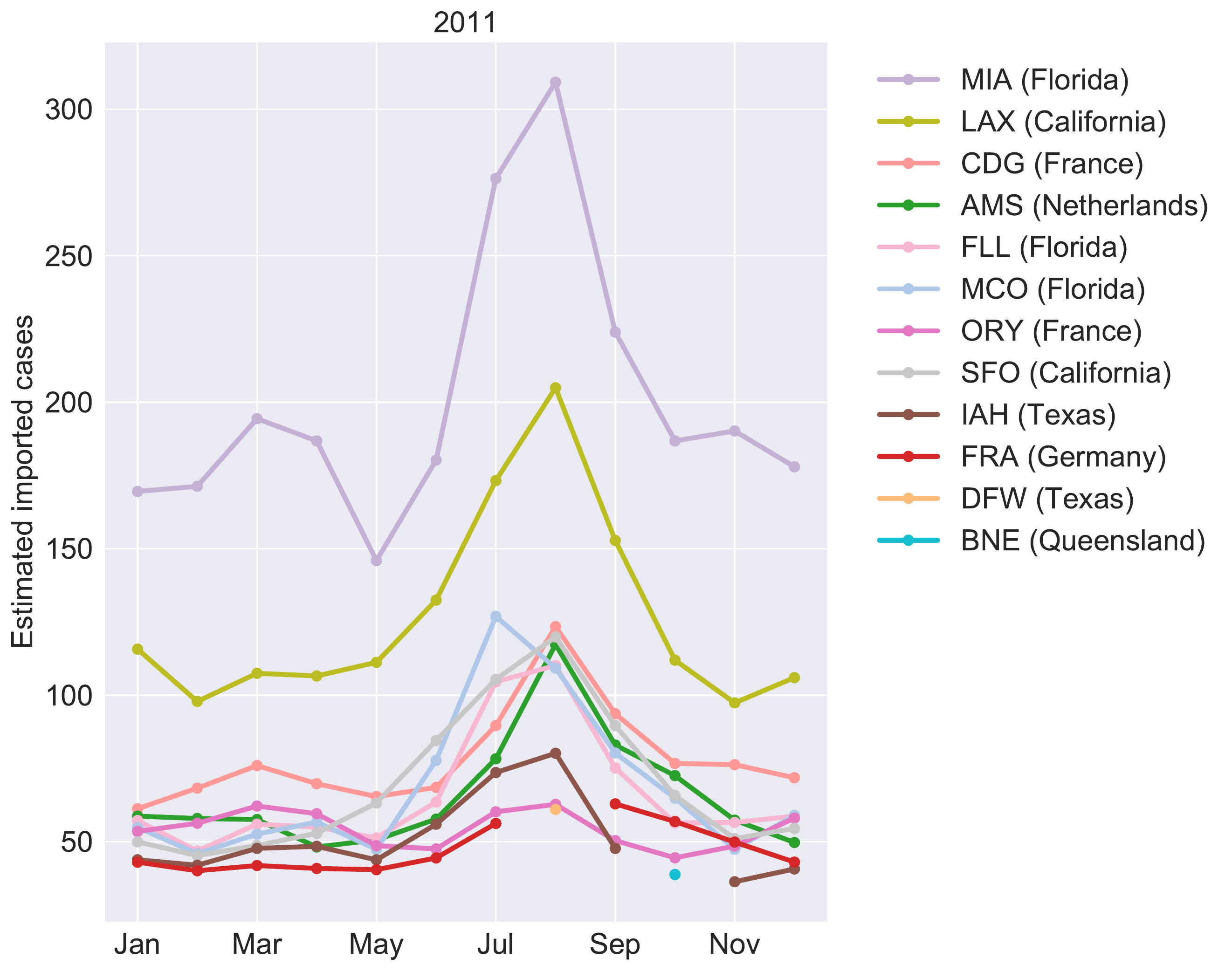}
	\caption{{\bf Predicted monthly dengue importations by airport for 2011.} \normalfont The number of predicted imported dengue infections for the top ten airports in non-endemic countries/states with vector presence for each month in 2011. A break in a line indicates that the corresponding airport was not amongst the top ten during the respective month. Airports are abbreviated using the corresponding IATA code. A full list of abbreviations can be found in the supplementary material (see Table S1)}
	\label{fig:Top10_2011}
\end{figure}

\begin{figure}[h]
	\centering
	\includegraphics[width=\textwidth]{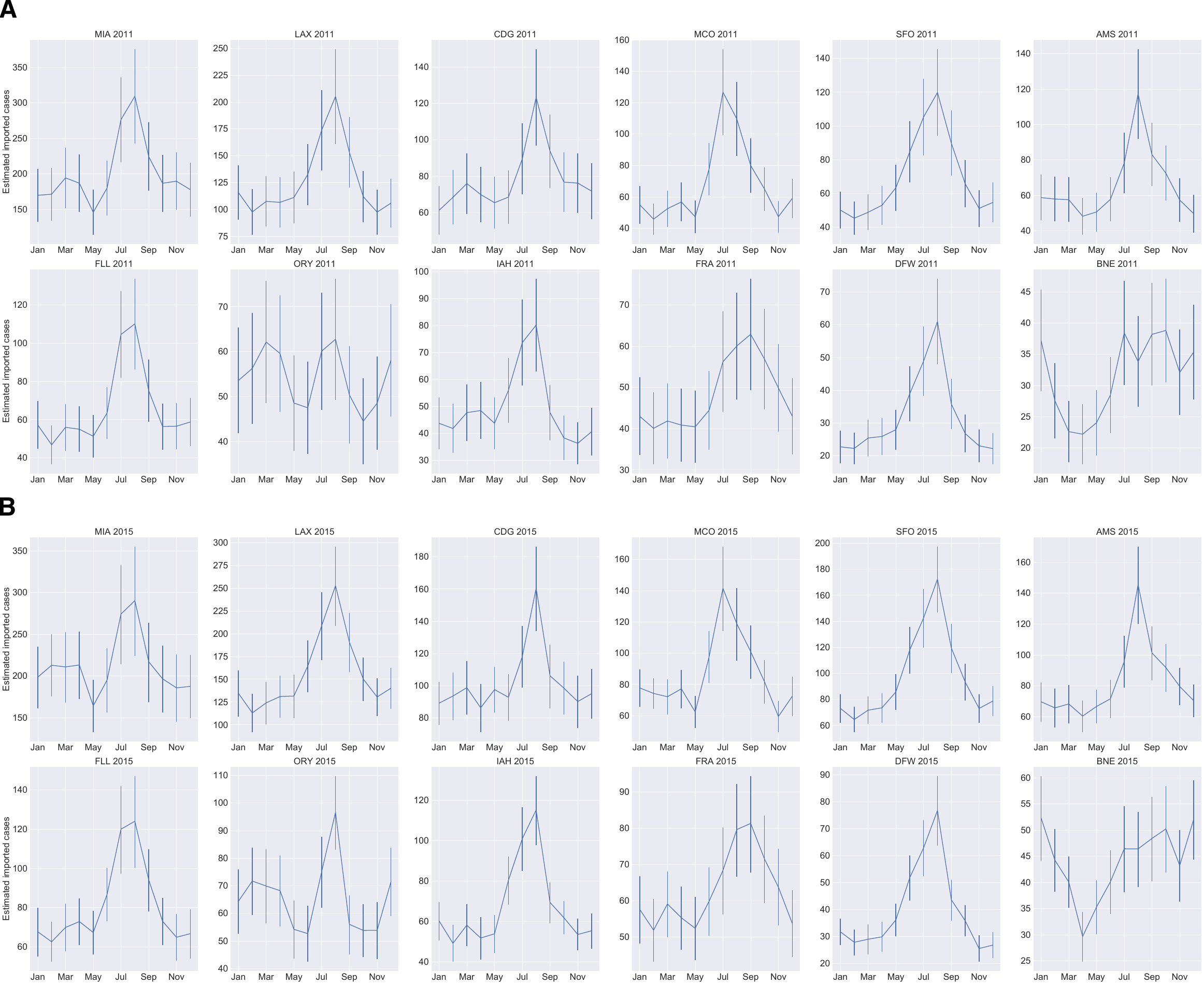}
	\caption{{\bf Predicted monthly dengue importations by airport} \normalfont The number of predicted imported dengue infections for the top ten airports in non-endemic countries/states with vector presence for each month in ({\bf A}) 2011 and ({\bf B}) 2015. The error bars correspond to $\pm 1$ standard deviation. Airports are abbreviated using the corresponding IATA code.}
	\label{fig:Top10_CI}
\end{figure}

\begin{figure}[h]
	\centering
	\includegraphics[width=\textwidth]{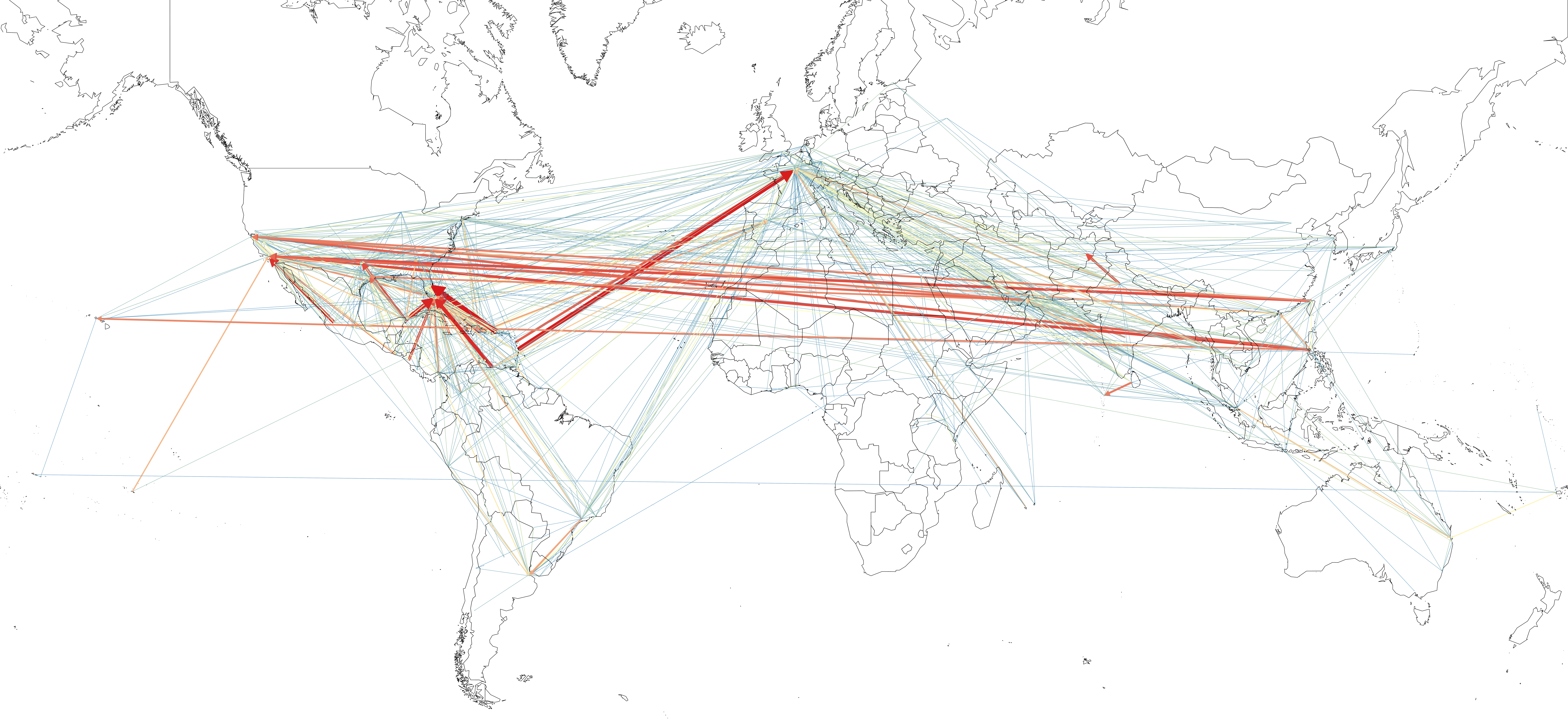}
	\caption{{\bf Dengue-infected passengers who continue to travel to non-endemic countries/states with vector presence for every route in the air transportation network} \normalfont This map corresponds to August 2015. The thickness as well as the colour of an edge represent the number of infected people travelling along the corresponding route. Blue represents relatively lower numbers of infected people, red represents relatively higher numbers of infected travellers and yellow represents the mid range.}
	\label{fig:importation_routes_Aug_2015}
\end{figure}

\vspace*{5mm}
\begin{figure}[h]
	\centering
	\includegraphics[width=0.8\textwidth]{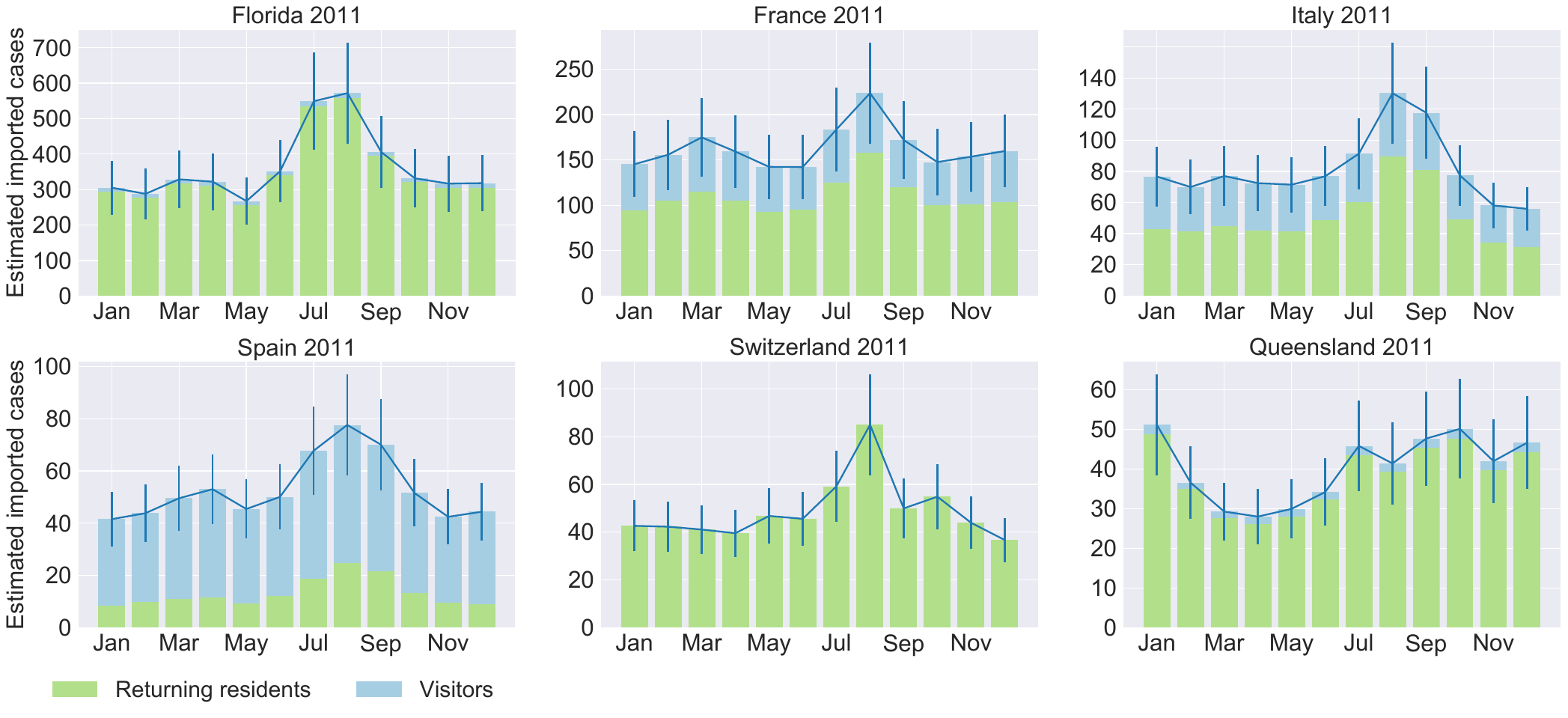}
	\caption{{\bf Predicted dengue infections imported by returning residents and visitors in 2011} \normalfont Here we show the results for non-endemic countries/states with vector presence with the highest number of predicted imported dengue cases in 2011. The bars are stacked to distinguish between returning residents (green) and visitors (blue). The blue solid line corresponds to the total number of imported cases. The error bars correspond to the model's coefficient of variation (see Material and methods).}
	\label{fig:visitorsResidents2011}
\end{figure}

\begin{figure}[h]
	\centering
	\includegraphics[width=\textwidth]{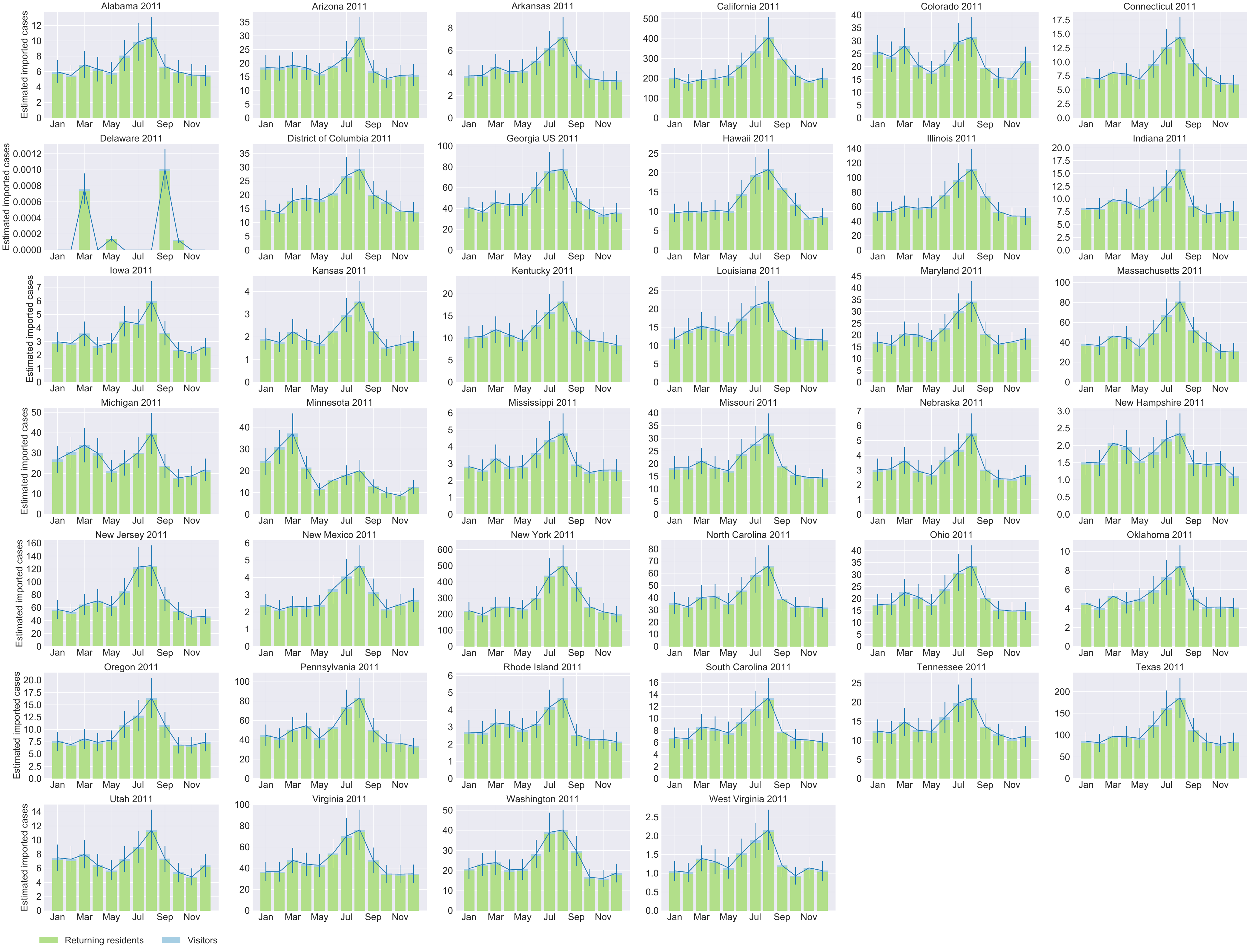}
	\caption{{\bf Predicted imported dengue infections for returning residents and visitors for US states in 2011.} \normalfont The bars are stacked to distinguish between returning residents (green) and visitors (blue). The blue solid line corresponds to the total number of imported cases. The error bars correspond to the model's coefficient of variation (13.49\%) that was inferred through Monte Carlo simulations.}
	\label{fig:US2011}
\end{figure}

\clearpage
\begin{figure}[h]
	\centering
	\includegraphics[width=\textwidth]{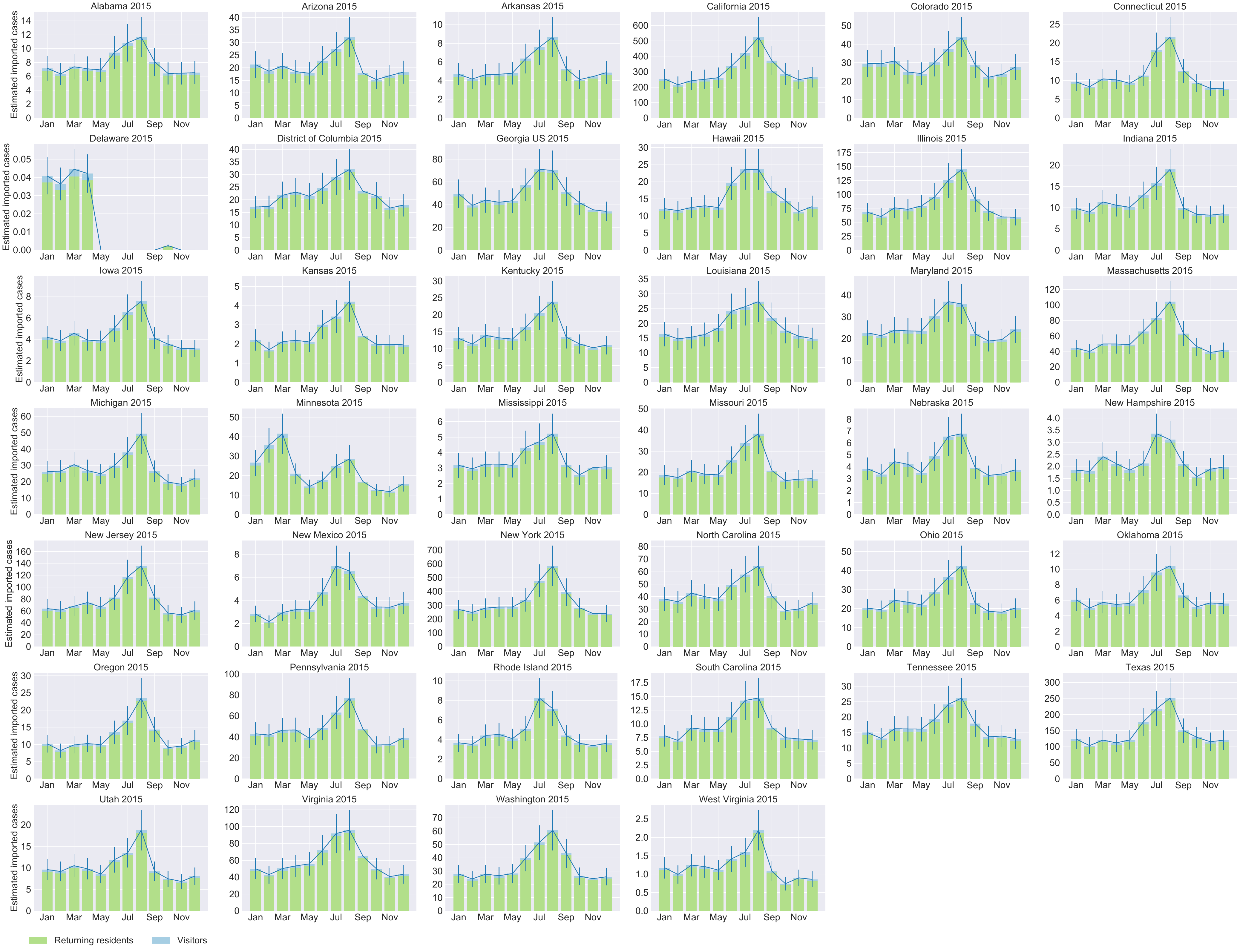}
	\caption{{\bf Predicted imported dengue infections for returning residents and visitors for US states in 2015.} \normalfont The bars are stacked to distinguish between returning residents (green) and visitors (blue). The blue solid line corresponds to the total number of imported cases. The error bars correspond to the model's coefficient of variation (13.49\%) that was inferred through Monte Carlo simulations.}
	\label{fig:US2015}
\end{figure}

\clearpage
\begin{figure}[h]
	\centering
	\includegraphics[width=\textwidth]{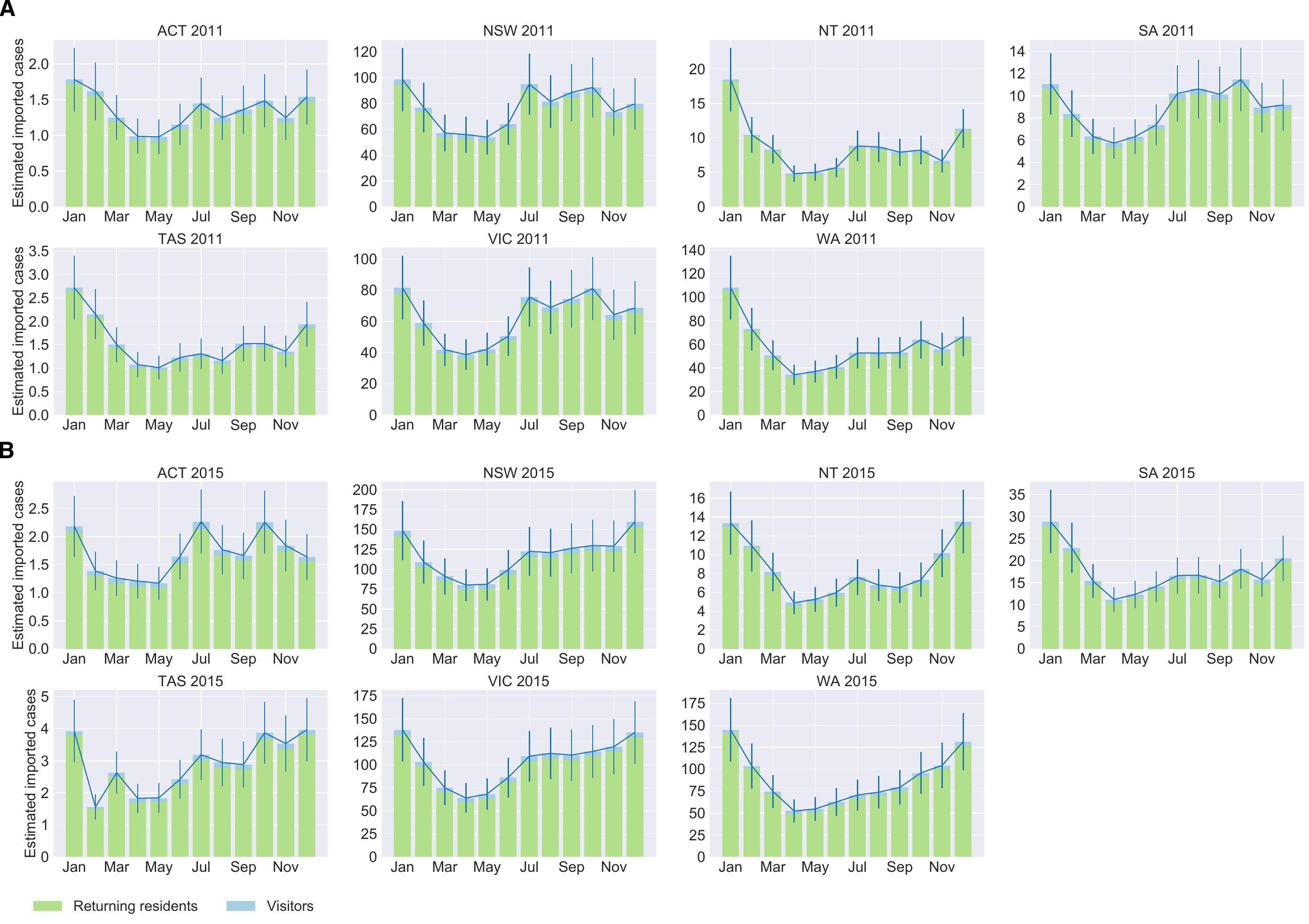}
	\caption{{\bf Predicted imported dengue infections for returning residents and visitors for Australian states.} \normalfont ACT: Australian Capital Territory, NSW: New South Wales, NT: Northern Territory, SA: South Australia, TAS: Tasmania, VIC: Victoria, WA: Western Australia. The bars are stacked to distinguish between returning residents (green) and visitors (blue). The blue solid line corresponds to the total number of imported cases. The error bars correspond to the model's coefficient of variation (13.49\%) that was inferred through Monte Carlo simulations. }
	\label{fig:AUS}
\end{figure}

\clearpage
\begin{figure}[h]
	\centering
	\includegraphics[width=\textwidth]{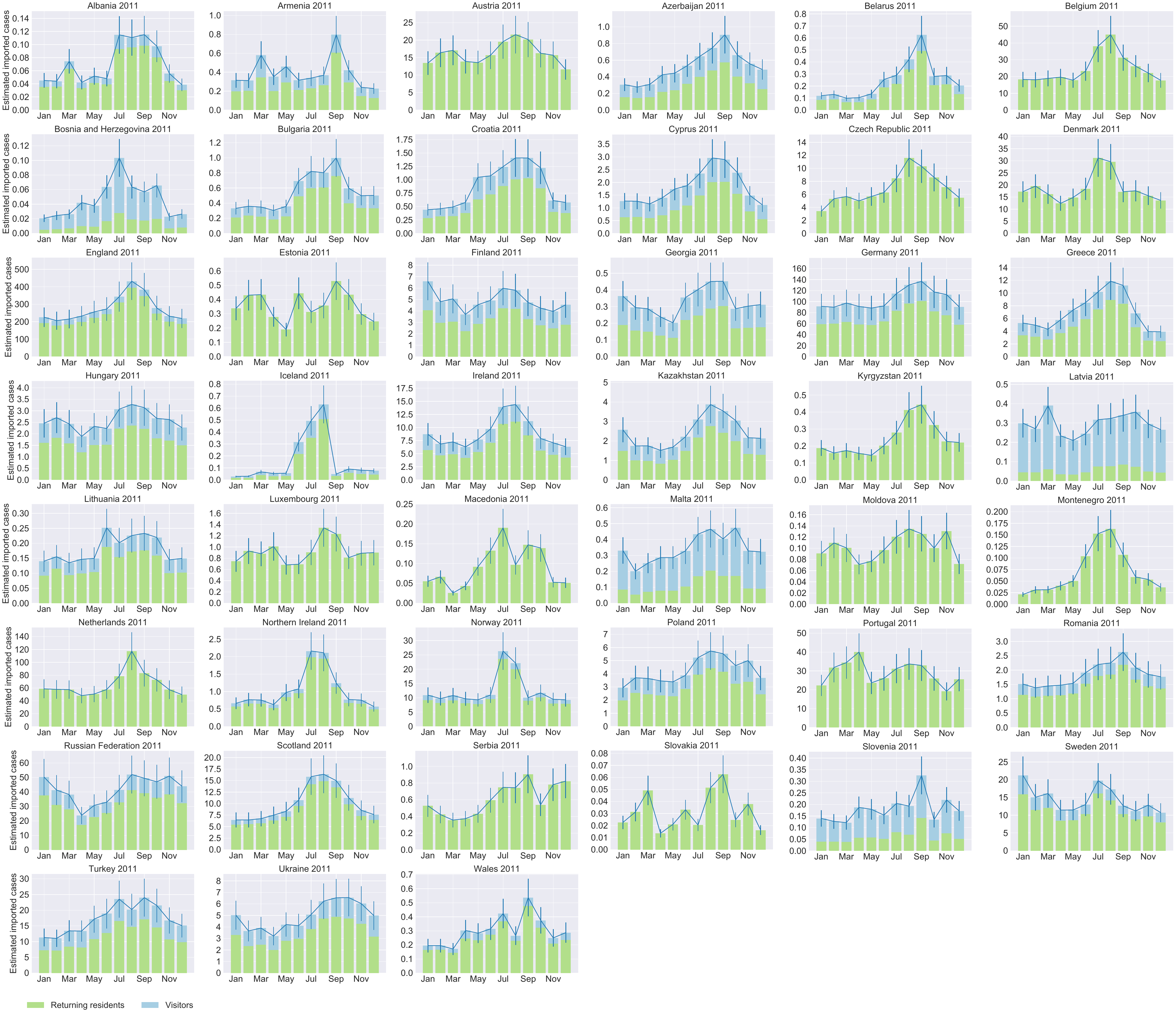}
	\caption{{\bf Predicted imported dengue infections for returning residents and visitors for European countries in 2011.} \normalfont The bars are stacked to distinguish between returning residents (green) and visitors (blue). The blue solid line corresponds to the total number of imported cases. The error bars correspond to the model's coefficient of variation (13.49\%) that was inferred through Monte Carlo simulations.}
	\label{fig:Europe2011}
\end{figure}

\clearpage
\begin{figure}[h]
	\centering
	\includegraphics[width=\textwidth]{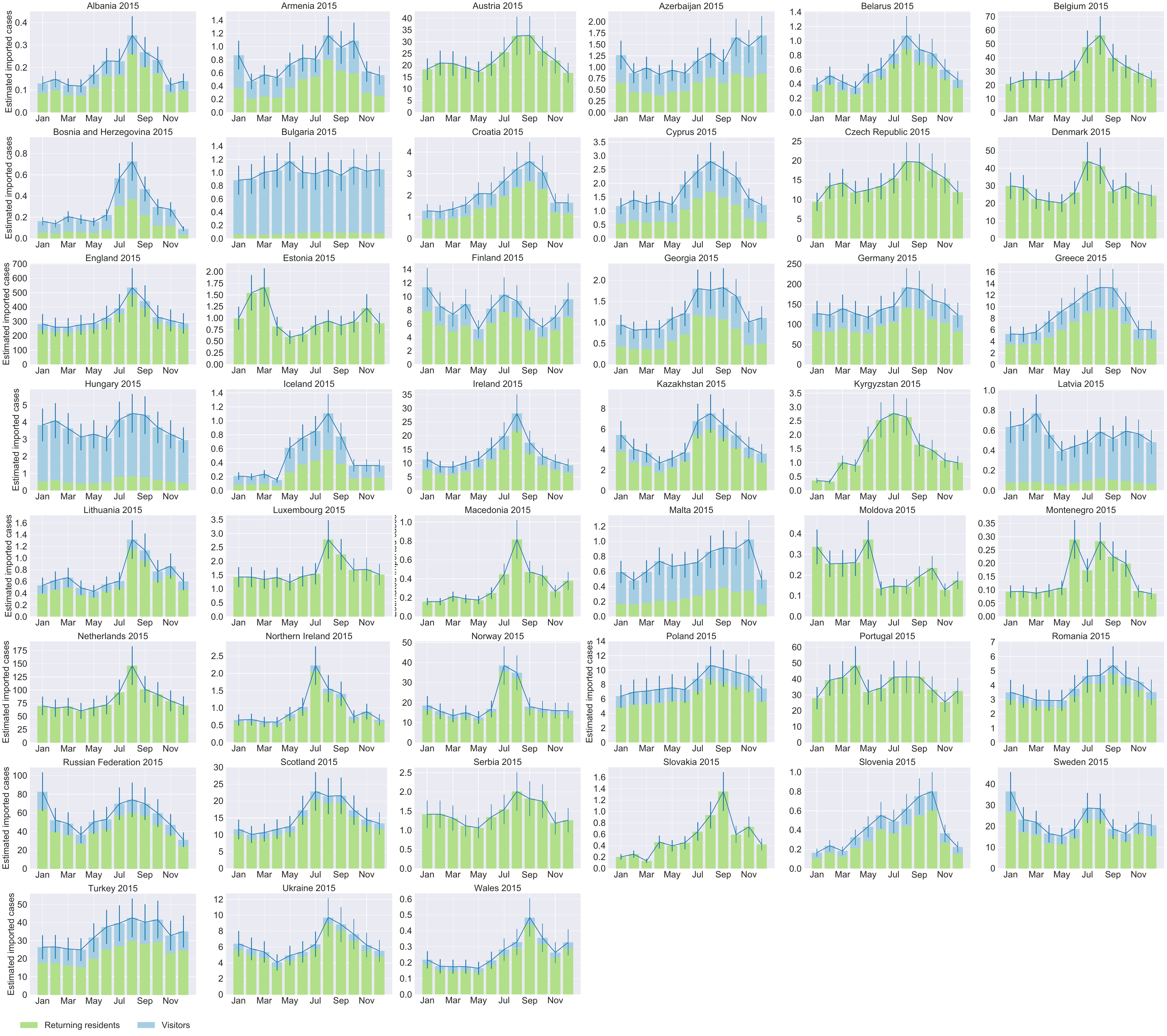}
	\caption{{\bf Predicted imported dengue infections for returning residents and visitors for European countries in 2015.} \normalfont The bars are stacked to distinguish between returning residents (green) and visitors (blue). The blue solid line corresponds to the total number of imported cases. The error bars correspond to the model's coefficient of variation (13.49\%) that was inferred through Monte Carlo simulations.}
	\label{fig:Europe2015}
\end{figure}

\clearpage
\begin{figure}[h]
	\centering
	\includegraphics[width=\textwidth]{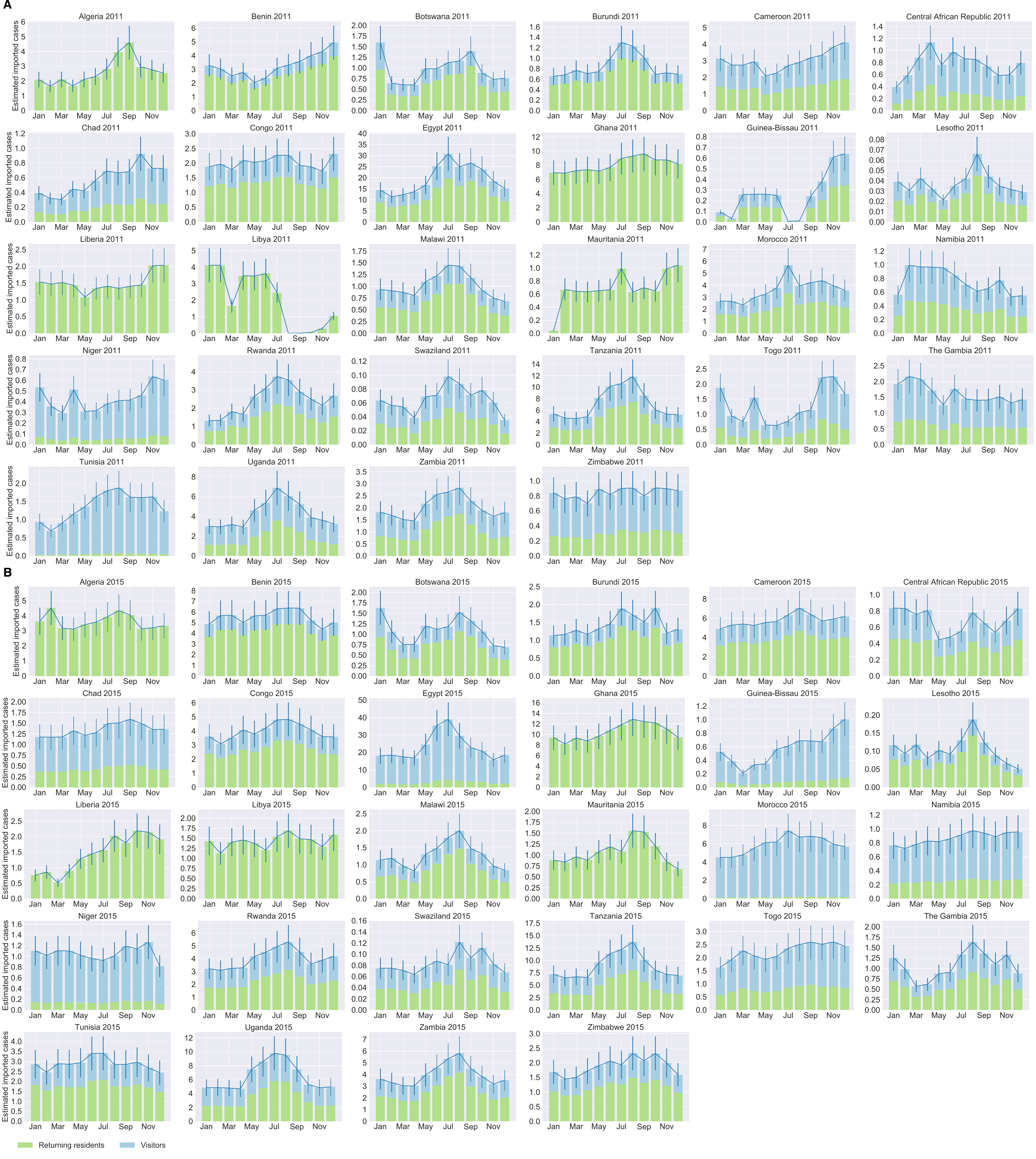}
	\caption{{\bf Predicted imported dengue infections for returning residents and visitors for non-endemic African countries.} \normalfont The bars are stacked to distinguish between returning residents (green) and visitors (blue). The blue solid line corresponds to the total number of imported cases. The error bars correspond to the model's coefficient of variation (13.49\%) that was inferred through Monte Carlo simulations.}
	\label{fig:Africa}
\end{figure}

\clearpage
\begin{figure}[h]
	\centering
	\includegraphics[width=\textwidth]{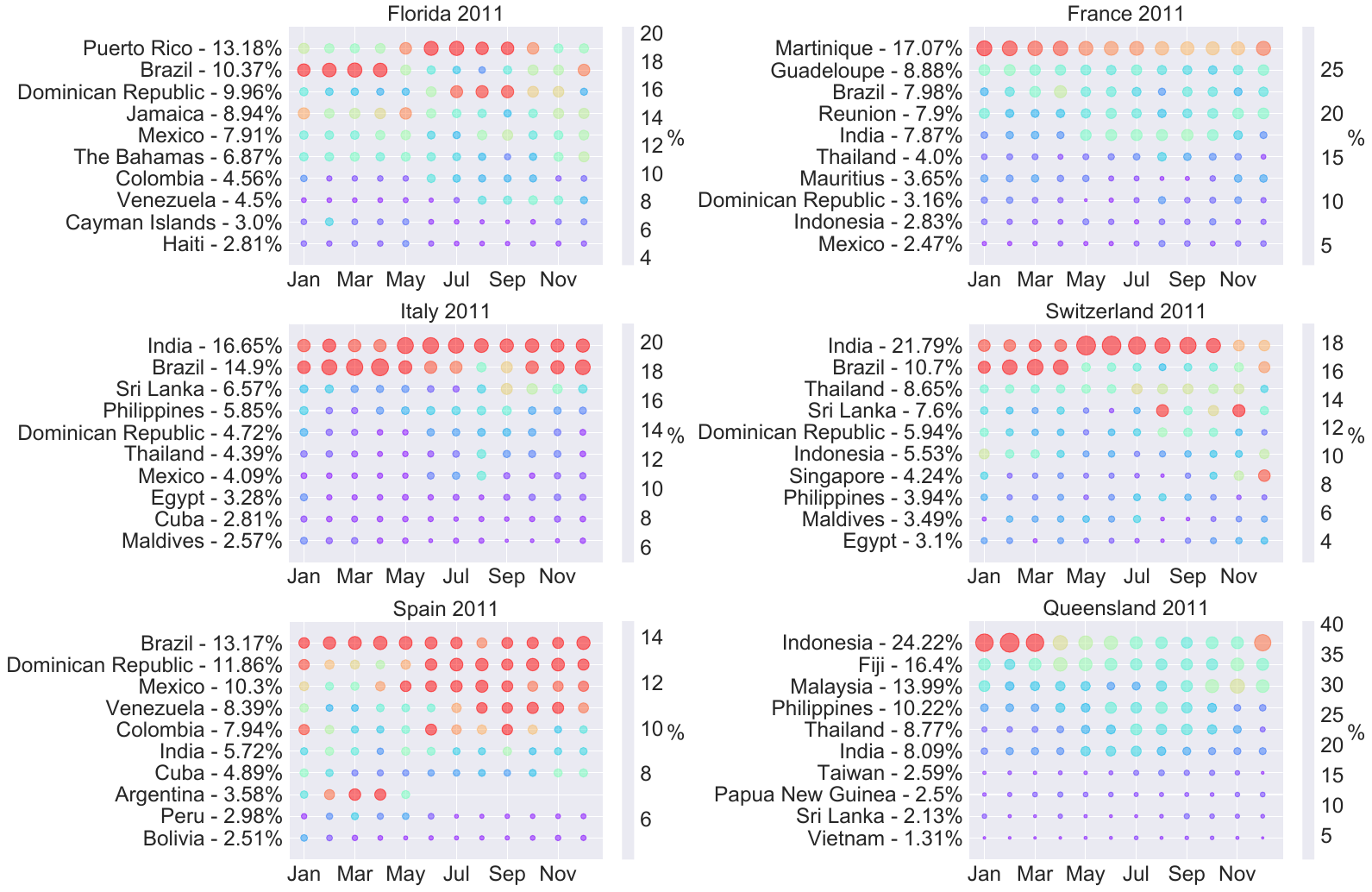}
	\caption{{\bf Predicted percentage contribution of dengue importations by country of acquisition in 2011.} \normalfont The predicted percentage contribution by source country and month in 2011. The size and colour of the circles indicate the percentage contribution of the corresponding country to the total number of imported cases. The $y$-labels indicate the yearly percentage contribution of the corresponding source country.}
	\label{fig:percentage_acquisition_2011}
\end{figure}

\clearpage
\begin{figure}[h]
	\centering
	\includegraphics[width=\textwidth]{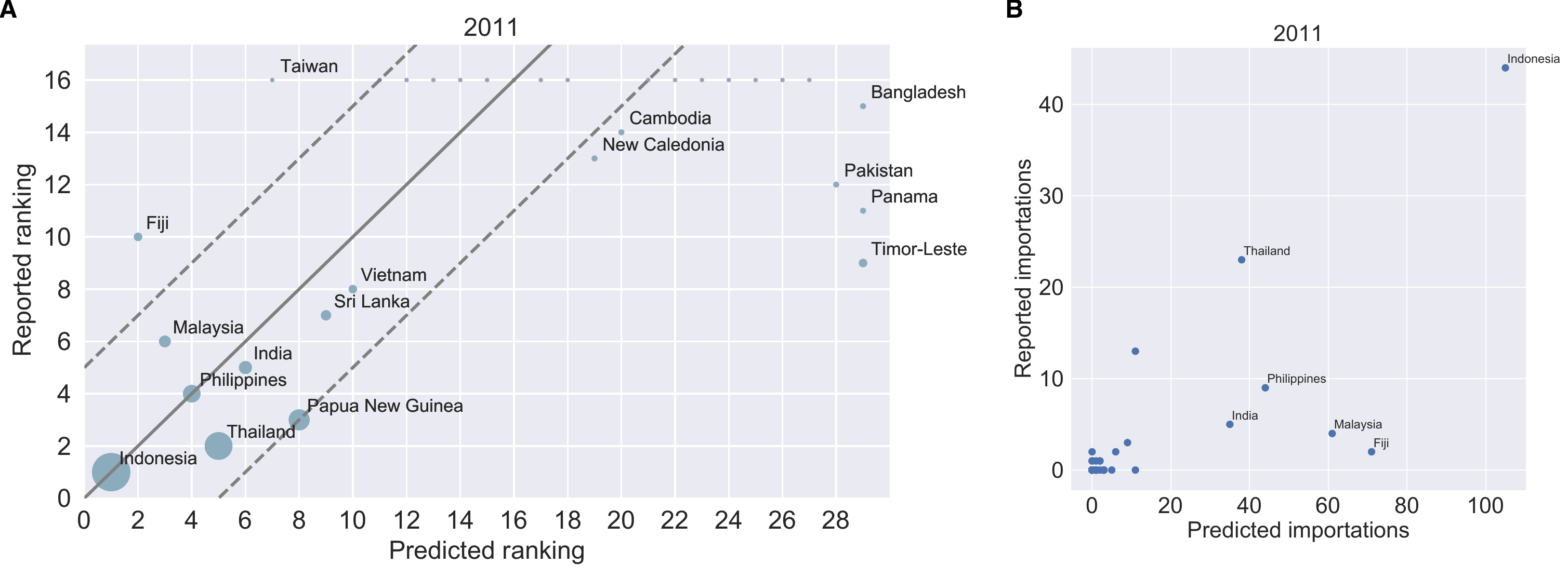}
	\caption{{\bf Rank-based validation and correlation between reported and predicted imported cases for Queensland in 2011.} \normalfont ({\bf A}) Countries are ranked by the total number of predicted and reported imported dengue cases. The reported ranking is then plotted against the predicted ranking. Countries that were ranked by the model, but did not appear in the dataset receive a rank of $i+1$, were $i$ is the number of unique importation sources according to the dengue case data. Similarly, countries that appeared in the data and were not ranked by the model receive a rank of $i+1$. For circles that lie on the $x=y$ line (grey solid line) the predicted and reported rankings are equal. Circles that lie between the two dashed lines correspond to countries with a difference in ranking that is less than or equal to five. The circle areas are scaled proportionally to the number of reported cases that were imported from the corresponding country. Spearman's rank correlation coefficient between the absolute numbers of reported and predicted importations is equal to 0.58. ({\bf B}) The absolute number of reported dengue importations are plotted against the absolute number of predicted importations.}
	\label{fig:rankingQLD2011}
\end{figure}

\begin{figure}[h]
	\centering
	\includegraphics[width=0.8\textwidth]{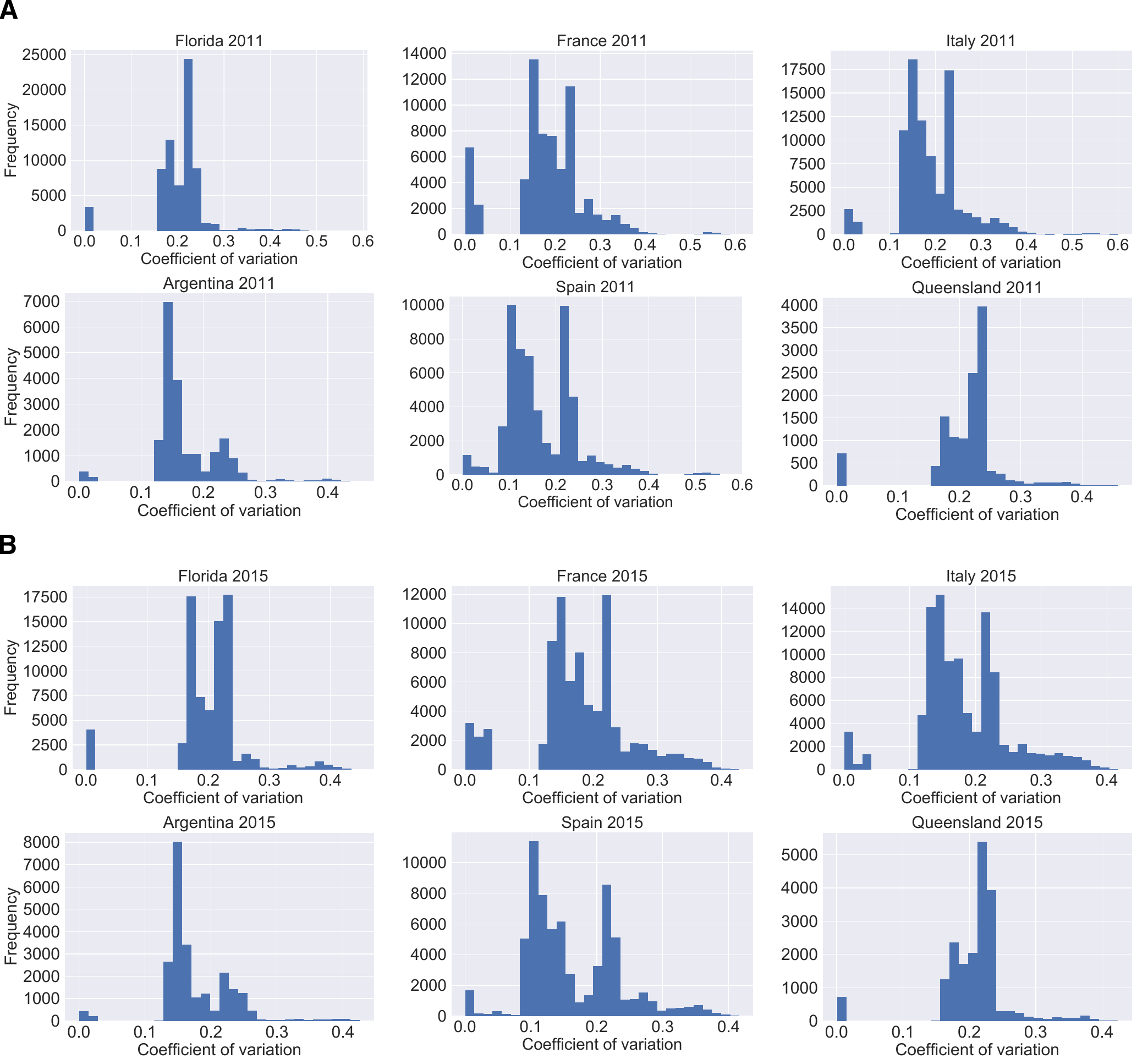}
	\caption{{\bf The distribution of the coefficient of variation for several destinations.} \normalfont ({\bf A}) Distributions for 2011. ({\bf B}) Distributions for 2015.}
	\label{fig:cv}
\end{figure}

\begin{figure}[h]
	\centering
	\includegraphics[width=\textwidth]{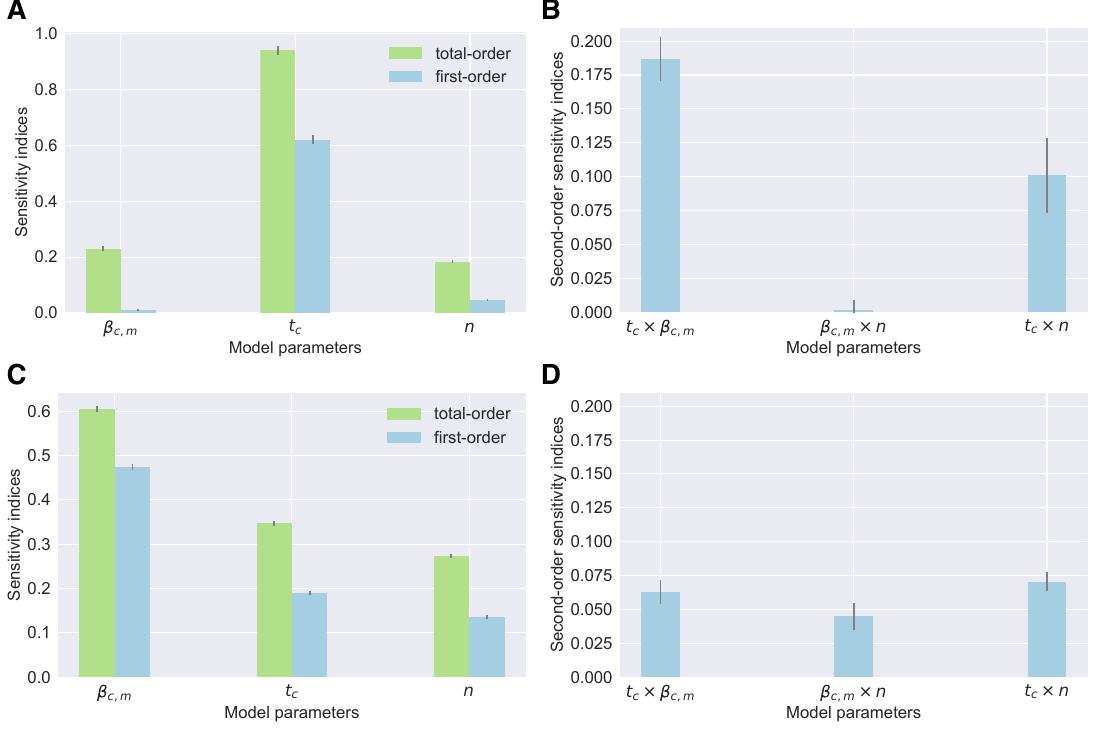}
	\caption{{\bf Sobol's sensitivity analysis of the model's parameters.} \normalfont Parameter $\beta_{c,m}$ denotes the daily dengue incidence rate of country $c$ during month $m$, parameter $t_{c}$ denotes the number of days a traveller who arrives at a given airport has spent in country $c$ and parameter $n$ denotes the sum of the intrinsic incubation period and the infectious period in humans. ({\bf A}) The first-order and total-order indices for the parameter ranges as shown in Table 1 of the main manuscript. The indices indicate that $t_c$ is the most important model parameter. ({\bf B}) The second-order indices for the parameter ranges as shown in Table 1 of the main manuscript. There is significant interactions between parameters $t_c$ and $\beta_{c,m}$ and between parameters $t_c$ and $n$. ({\bf C}) The first-order and total-order indices for a shorter range of value ([1, 30] days) for parameter $t_c$. In this case $\beta_{c,m}$ is the most important parameter. ({\bf D}) The second-order indices for a shorter range of value ([1, 30] days) for parameter $t_c$. There is still significant interaction between parameters $t_c$ and $\beta_{c,m}$ and between parameters $t_c$ and $n$.}
	\label{fig:sensitivity}
\end{figure}

	%----------------------------------------------------------------------------------------
	%	REFERENCE LIST
	%----------------------------------------------------------------------------------------
	
	%\bibliographystyle{abbrv} 
	%\bibliography{DengueImportation}
	
	%----------------------------------------------------------------------------------------
	
\end{document}